\documentclass[aps,prb,reprint,groupedaddress,twocolumn]{revtex4-1}
\usepackage{graphicx}
\usepackage{gensymb}
\usepackage[final]{changes}

\begin{document}


\title{Vector magnetometry based on $\mathrm{S=3/2}$ electronic spins}


\author{Sang-Yun Lee}
\email{s.lee@physik.uni-stuttgart.de}
\author{Matthias Niethammer}
\author{J$\mathrm{\ddot{o}}$rg Wrachtrup}
\affiliation{3rd institute of Physics, University of Stuttgart, Pfaffenwaldring 57, 70569 Stuttgart, Germany}


\date{\today}
\definechangesauthor[name={Sang-Yun Lee},color=red]{SYL}

\begin{abstract}
Electronic spin systems with $\mathrm{S>1/2}$ provide an efficient method for DC vector magnetometry, since the conventional electron spin resonance spectra at a given magnetic field reflect not only the field strength but also orientation in the presence of strong spin-spin interactions. S=1 spins, e.g. the nitrogen-vacancy centers in diamond, have been intensively investigated for such a purpose. In this report, we compare S=1 and S=3/2 spins, and discuss how one can apply general principles for the use of high spin systems as a vector magnetometer to the S=3/2 spin systems. We find analytical solutions which allow a reconstruction of the magnetic field strength and polar angle using the observed resonance transitions if an uniaxial symmetry exists for the spin-spin interaction as in S=1 systems. We also find that an ambiguity of determining the field parameters may arise due to the unique properties of S=3/2 systems, and present solutions for it utilizing additional transitions in the low-field region. The electronic spins of the silicon vacancy in silicon carbide will be introduced as a model for the S=3/2 DC vector magnetometer and the practical usage of it, including the magic-angle spinning type method, will be presented too.
\end{abstract}

\maketitle

\section{Introduction\label{section:intro}}
Electronic spins in highly localized defects, such as the nitrogen-vacancy (NV) centers in diamond~\cite{Doherty2013,Schirhagl2014} and vacancy related defects in silicon carbide (SiC)~\cite{Sorman2000,Janzen2009,Widmann2015,Christle2015,Weber2010a,Koehl2011a,Baranov2011,Mizuochi2002,Kraus2014,Szasz2015,Gali2010,Son2006a}, may experience a strong spin-spin interaction, e.g., a dipole-dipole interaction, which results in the so-called zero-field splitting (ZFS), partially (or completely) lifting degeneracy of energy eigenstates at zero magnetic field~\cite{Stevenson1984,Atherton1993}. If this interaction is strong enough,
the eigenvalues of the spin Hamiltonian show a strong dependence on the orientation of the applied magnetic field. Such dependence causes a non linear shift of resonance transitions in electron spin resonance (ESR) spectra. Thus, the information about the applied external magnetic field can be extracted from ESR spectra provided the ZFS is known.

One well-known example is the NV center in diamond. Its application to DC field vector magnetometry has been reported and well understood in the field strength from sub-$\mathrm{\mu T}$ to a few tenth T~\cite{Balasubramanian2008,Steinert2010,Clevenson2015,Taylor2008}. The NV center has a triplet ground state of $\mathrm{S=1}$ and when shifts of the ESR transition at a given DC magnetic field are directly monitored in the frequency domain, typically a $\mathrm{\sim 0.1~mT}$ minimum detectable magnetic field is achieved~\cite{Degen2008}. This resolution is limited by the ESR linewidth which can be broadened by strong RF fields thus lowering the resolution. Lower RF power can be used to avoid power broadening, but the decreased signal strength requires a very long accumulation time. If time-domain experiments, e.g. a Ramsey fringe experiment, in which the magnetic field strength is imprinted in the phase of the superpositioned state, is conducted, a large signal strength can be maintained without power broadening, thus a sensitivity up to $\sim\mathrm{0.4~\mu T/\sqrt{Hz}}$, limited by the $\mathrm{T_{2}^{*}}$ of $\mathrm{\sim 1~\mu s}$, can be realized using a single NV center~\cite{Taylor2008,Schirhagl2014}. Further enhancement (below $\mathrm{1~nT/\sqrt{Hz}}$) is possible by using the NV center ensemble combined with the lock-in detection~\cite{Clevenson2015}. When the NV center is used for AC magnetic field sensing, spin echo type measurements can be used in which the long coherence time allows high sensitivity up to $\mathrm{\sim1~nT/\sqrt{Hz}}$ using a single NV~\cite{Balasubramanian2009a} and  $\mathrm{\sim0.9~pT/\sqrt{Hz}}$ using an NV ensemble~\cite{2014arXiv1411.6553W}.

Higher spin systems ($\mathrm{S>1}$) can also be used as a vector magnetometer in a similar way. For example, the silicon vacancy ($\mathrm{V_{Si}}$) in silicon carbide (SiC) is known to posses\added{s} a quartet manifold of $\mathrm{S=3/2}$ in its electronic ground state~\cite{Mizuochi2003,Isoya2008,Mizuochi2005prb}. Because its ESR signal can be detected at ambient condition~\cite{Janzen2009,Soltamov2012,Simin2015, Kraus2014a, Janzen2009,Kraus2014} even from a single defect~\cite{Widmann2015} and the ZFS is in a range around a few mT depending on the polytype of SiC~\cite{Janzen2009,Sorman2000}, its application as a DC magnetometer has been suggested~\cite{Kraus2014a,Simin2015}. Note that Simin \textit{et al}., have recently shown an experimental application of $\mathrm{V_{Si}}$ in SiC as a sub-mT DC magnetometer \replaced{based on approximated solutions for the spin Hamiltonian at weak magnetic fields}{by a numerical solutions of the spin Hamiltonian at a given magnetic field}~\cite{Simin2015}.

\begin{figure}
\includegraphics[width=1\columnwidth]{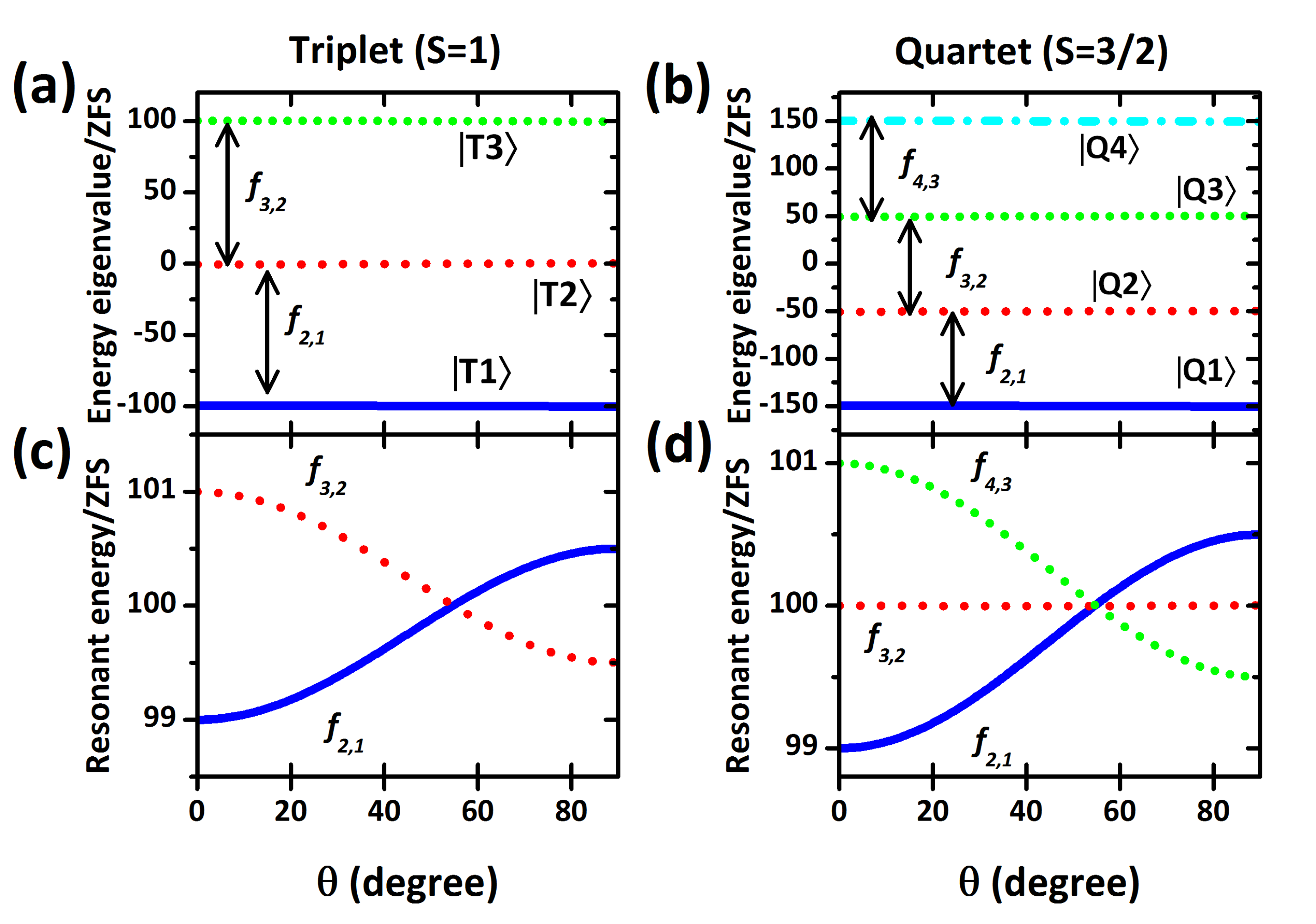}

\caption{\label{fig:TripletvsQuartet}(Color online) Energy eigenvalues and ESR transition energies of high spin systems as a function of orientation at \added{a }high magnetic field \added{$g\mu_{B}$}$\mathrm{B_{0}}=100\times \mathrm{ZFS}$. (a) and (b) are the eigenvalues of \added{the} spin triplet (S=1) and quartet (S=3/2) state calculated numerically from Eq.(\ref{eq:hamiltonian}), respectively. For numerical calculation, $D,E>0$, and an uniaxial symmetry $E\ll D$ are assumed thus only the polar angle, $\theta$ dependence is shown. Eigenstates are labeled in ascending order of corresponding energy eigenvalues. The most dominant transitions are indicated by the solid arrows with labels $f_{i,j}$ indicating \added{a} transition between $\mathrm{|i\rangle}$ and $\mathrm{|j\rangle}$. They are shown in (c) and (d) for the spin triplet and quartet states, respectively. All the energy values are normalized by ZFS.}


\end{figure}


In a spin system with a spin quantum number $\mathrm{S}$, the strength and orientation of the applied magnetic field vector $\mathbf{B_{0}}$ determines the Zeeman splitting, thus one should experimentally obtain the Zeeman splitting to get information about $\mathbf{B_{0}}$. For $\mathrm{S=1/2}$, the Zeeman splitting is calculated from an observed single resonant transition energy $h \nu = g \mu_{B} B_{0}$ where g is the Land\'{e} g-factor, $\mu_{B}$ is the Bohr magneton, and $h$ is \replaced{Planck's}{the planck} constant. The information for the orientation can be extracted only if $g$ is anisotropic. In high spin systems, the orientation related terms remain in the eigenvalue equation which result in the orientation dependent shift of ESR spectra\added{, which cannot be explained by $g \mu_{B} B_{0}$}. It is, therefore, mandatory to reconstruct the energy eigenstates using the observed resonant energies. Because there exist 2S+1 eigenstates, 2S resonant transition energies should be experimentally determined. For example, when \added{the applied magnetic field strength is much larger than the ZFS, i.e., $g\mu_{B}$}$|\mathbf{B_{0}}|\gg \mathrm{ZFS}$, at least two transition energies should be known for $\mathrm{S=1}$ while three values are necessary for $\mathrm{S=3/2}$ as explained in Fig.\ref{fig:TripletvsQuartet}. In this high field range, the two transitions for $\mathrm{S=1}$ and two out of the three transitions for $\mathrm{S=3/2}$ cross each other as shown in Figs.\ref{fig:TripletvsQuartet}(c) and 1(d). This leads to ambiguity \replaced{i}{o}n determining \replaced{which observed ESR peak corresponds to which transition energy}{the transition energies} experimentally. \deleted{However, this problem can be avoided if one can find analytical solutions for the given $\mathrm{\mathbf{B_{0}}}$ vector in terms of the resonant transition energies which are invariant under switching those two transition energies~\cite{Balasubramanian2008}. }In this report, we discuss how this ambiguity can be removed and thus show how to use \added{the} $\mathrm{S=3/2}$ system for vector magnetometry. For this, we will provide analytical solutions for the given $\mathrm{\mathbf{B_{0}}}$ vector as a function of the resonant transition energies. We will present $\mathrm{V_{Si}}$ spins in SiC as a model system and also a novel magnetomet\deleted{e}ry scheme using the magic angle. The discussion in this report is also applicable to other $\mathrm{S=3/2}$ systems which have been found in fullerene~\cite{Morton2005JCP,Knapp1998,Harneit2002pra,Benjamin2006JP,MizuochiJCP1999},
organic molecules~\cite{Teki2001JACS,Kothe1980cpl,Teki2008}, Ni impurities in diamond~\cite{Isoya1990}, and 
calcium oxide crystals~\cite{vanLeeuwen19860rb}.


\section{Vector magnetometry based on S=3/2 spins\label{section:eigenvalue}}
\begin{figure}
\includegraphics[width=1\columnwidth]{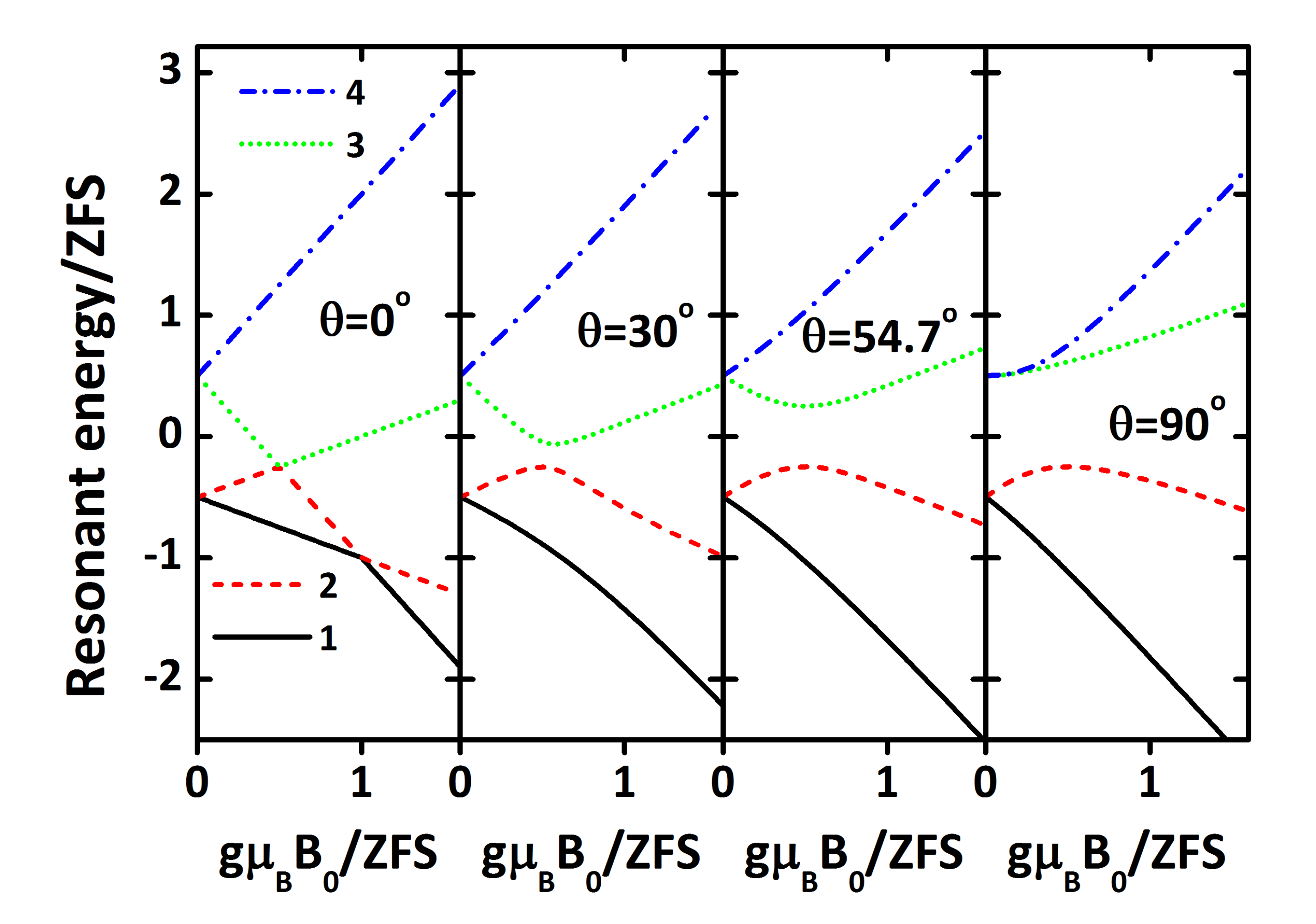}
\caption{\label{fig:eigenvaluesvsB0}(Color online) $B_{0}$ dependenc\replaced{e}{ies} of the energy eigenvalues of a spin quartet state at various magnetic field orientations. Eigenstates are labeled in ascending order of the corresponding energy eigenvalues. As $B_{0}\rightarrow\infty$ at $\theta = 0\,\degree$, $\mathrm{|4\rangle \rightarrow |m_{S}=+3/2\rangle}$, $\mathrm{|3\rangle \rightarrow |m_{S}=+1/2\rangle}$, $|\mathrm{2\rangle \rightarrow |m_{S}=-1/2\rangle}$, and $|\mathrm{1\rangle \rightarrow |m_{S}=-3/2\rangle}$.
}
\end{figure}

In order to derive formulas for $B_{0}\equiv |\mathrm{\mathbf{B_{0}}}|$ and its orientation expressed by only three transition energies for $\mathrm{S=3/2}$, we will first construct the electronic spin Hamiltonian consisting of the ZFS and Zeeman term. The ZFS in high spin systems can be described by the dipole-dipole interaction term in the spin Hamiltonian, $\mathrm{\mathbf{S \cdot D \cdot S}}$ where $\mathrm{\mathbf{D}}$ is the dipole-dipole coupling tensor. For simplicity, we assume an isotropic Land\'{e} $g$-factor. \replaced{Therefore}{Then} in the principal axis system of $\mathrm{\mathbf{D}}$, in which the z-axis is set to the symmetry axis, the electronic spin Hamiltonian at  $\mathrm{\mathbf{B_{0}}}$, is
\begin{equation}\label{eq:hamiltonian}
H=g_{} \mu_{B} \mathbf{B_{0}}\cdot \mathbf{S} +D\{S_{z}^{2}-S(S+1)/3\}+E(S_{+}^{2}+S_{-}^{2} )/2\added{,}
\end{equation}
where $E$ and $D$ are the ZFS parameters, assumed to be positive, and $E\ll D$ if an uniaxial symmetry exists.
For S=3/2, the eigenvalue equation from Eq.(\ref{eq:hamiltonian}) is, in the polar coordinate system,
\begin{widetext}
\begin{eqnarray}\label{eq:eigenvalue_equation}
&&\lambda^{4}-(2D^{2}+6E^{2}+\frac{5}{2}\beta_{0}^{2})\lambda^{2}-2\beta_{0}^{2}\{D(3 \cos ^{2} \theta -1)+3E \sin ^{2}\theta \cos 2\phi \} \lambda \nonumber \\
&&+\frac{9}{16}\beta_{0}^{4}+D^{4} -\frac{1}{2}D^{2}\beta_{0}^{2}-D^{2}\beta_{0}^{2}(3 \cos ^{2}\theta -1)+3E^{2}(3E^{2}+2D^{2})\nonumber \\
&&+E \beta_{0}^{2} (6D \sin ^{2} \theta \cos 2 \phi +\frac{9}{2} E \cos 2\theta     )=0\added{,}
\end{eqnarray}
\end{widetext}
where \replaced{$\beta_{0} \equiv g \mu_{B} B_{0}$}{$g$ and $\mu_{B}$ are omitted for convenience}. The numerically calculated eigenvalues at various orientations are shown in Fig.\ref{fig:eigenvaluesvsB0}. When $\mathbf{B_{0}}$ is either parallel or perpendicular to the symmetry axis, the closed form solutions for each eigenvalue\deleted{s} can be found as~\cite{Atherton1993}
\begin{widetext}
\begin{eqnarray}\label{eq:eigenvalueformula}
\lambda&=& \frac{1}{2} \beta_{0} \pm \sqrt{(D+\beta_{0})^{2}+3E^{2}} \,\,\,\mathrm{or} \,\,\,  -\frac{1}{2}\beta_{0}\pm \sqrt{(D-\beta_{0})^{2}+3E^{2}}    \,\,\,\mathrm{for}\,\,\,\mathbf{B_{0}} \| \mathrm{z-axis}\added{,} \nonumber \\
\lambda&=& \frac{1}{2}\beta_{0} \pm \sqrt{\beta_{0}^{2}+D^{2}+3E^{2}- (D-3E) \beta_{0}    } \,\,\,\mathrm{or} \nonumber \\
&&-\frac{1}{2}\beta_{0} \pm \sqrt{\beta_{0}^{2}+D^{2}+3E^{2}+ (D-3E) \beta_{0}    } \,\,\,\mathrm{for}\,\,\,\mathbf{B_{0}} \parallel \mathrm{x-axis}\added{,} 
\end{eqnarray}
\end{widetext}
which give\added{s} the eigenvalues at zero magnetic field, $\lambda_{B_{0}=0}=\pm \mathrm{ZFS}/2$ where $\mathrm{ZFS}\equiv 2 \sqrt{D^{2}+3E^{2}}$.
The eigenvalue equation for the general case can be expressed as,
\begin{equation}\label{generalsecular}
\sum_{n=0}^{2S+1} C_{n} \lambda^{n}=0.
\end{equation}
By plugging each eigenvalue $\lambda_{i}$ into Eq.(\ref{generalsecular}), one can obtain 2S+1 equations. The basic idea in order to find formulas for the $\mathrm{\mathbf{B_{0}}}$ vector expressed by the observed resonant energies, is to remove all $\lambda_{i}$ terms using the transition energy $f_{i,i-1}\equiv \lambda_{i}-\lambda_{i-1}$. Note that the energy eigenstates are not necessarily sorted with respect to the corresponding energy values. In other word, the indices can be randomly assigned to \replaced{the states}{each states}. Here, however, we keep the relation, $\lambda_{i}>\lambda_{i-1}$, for convenience. 
We follow this approach which has been frequently used for S=1 systems~\cite{Balasubramanian2008,DeGroot1960}. First, (2S-1) sets of three simultaneous equations are obtained by plugging $\lambda_{i}+f_{i+1,i}$, $\lambda_{i}$, and $\lambda_{i}-f_{i,i-1}$ (i=2,3,..2S) into Eq.(\ref{generalsecular}).
In each set, calculating
\begin{eqnarray}
&&\sum_{n=0}^{2S+1} \frac{C_{n} \{(\lambda_{i}+f_{i+1,i})^{n}-\lambda_{i}^{n} \}}{C_{2S+1}}=0 \,\, \mathrm{and} \nonumber \\
&&\sum_{n=0}^{2S+1} \frac{C_{n} \{(\lambda_{i}-f_{i,i-1})^{n}-\lambda_{i}^{n} \}}{C_{2S+1}}=0\added{,}
\end{eqnarray}
results in two new simultaneous equations
\begin{equation}
\sum_{n=0}^{2S} C_{i,n}^{\prime} \lambda_{i}^{n}=0 \,\, \mathrm{and} \,\, \sum_{n=0}^{2S} C_{i,n}^{\prime \prime} \lambda_{i}^{n}=0.
\end{equation}
Again, by subtracting one from each other divided by the coefficient of the highest order term of each equation, respectively, as below
\begin{equation}
\sum_{n=0}^{2S} \frac{C_{i,n}^{\prime} \lambda_{i}^{n}}{C_{i,2S}^{\prime}}-\frac{C_{i,n}^{\prime\prime} \lambda_{i}^{n}}{C_{i,2S}^{\prime\prime}}=0\added{,}
\end{equation}
we obtain a new equation for the eigenvalue of the energy eigenstate $|i\rangle$ in which the highest power is (2S-1),
\begin{equation}\label{newSecular}
\sum_{n=0}^{2S-1}C_{i,n}^{(2S-1)} \lambda_{i}^{n}=0\added{,}
\end{equation}
where i=2,3,..2S. By repeating this procedure until only one linear equation for a single eigenvalue remains, one can find a formula for an eigenvalue expressed in terms of resonant energies, which allows us to find expressions for all other eigenvalues again using $f_{i,i-1}$. We use this procedure to find solutions for S=3/2.

Following the procedure explained above, we obtain two equations for two energy eigenvalues, $\lambda_{2}$ and $\lambda_{3}$, expressed by only $f_{2,1}$ and $f_{3,2}$, and $f_{3,2}$ and $f_{4,3}$, respectively, and $\mathrm{B_{0}}$, $\theta$, and $\phi$ \replaced{which are present in both equations}{in common}. Then using $f_{3,2}\equiv \lambda_{3}-\lambda_{2}$ once again, we obtain formulas for each eigenvalue\deleted{s} expressed by only the resonant energies as below,
\begin{eqnarray}\label{eq:quarteteigenvalues}
\lambda_{1}&=&   -\frac{3}{4} f_{2,1}   -\frac{1}{2} f_{3,2}    -\frac{1}{4} f_{4,3},  \nonumber \\
\lambda_{2}&=&   \frac{1}{4} f_{2,1}   -\frac{1}{2} f_{3,2}    -\frac{1}{4} f_{4,3},  \nonumber \\
\lambda_{3}&=&   \frac{1}{4} f_{2,1}   +\frac{1}{2} f_{3,2}    -\frac{1}{4} f_{4,3},  \nonumber \\
\lambda_{4}&=&   \frac{1}{4} f_{2,1}   +\frac{1}{2} f_{3,2}    +\frac{3}{4} f_{4,3}.
\end{eqnarray}
And by plugging one of these, e.g. $\lambda_{2}$, back into one of the equations found in the preceding steps, we finally obtain formulas for $\beta_{0}^{2}$, and a new quantity related to $\theta$ and $\phi$, $\eta \equiv E\mathrm{(2 cos^{2}{\phi}\, sin^{2}{\theta}+cos^2{\theta})}+  D \mathrm{\cos^{2}{\theta}} $,
\begin{widetext}
\begin{eqnarray}
\beta_{0}^{2} &=& \frac{(\frac{\sqrt{3}}{2} f_{4,3} + f_{3,2}+ \frac{\sqrt{3}}{2} f_{2,1})^2+(1-\sqrt{3})(f_{4,3}+f_{2,1})f_{3,2}- f_{4,3} f_{2,1} -\mathrm{ZFS}^{2}}{5}\label{eq:B0}, \\
\eta &=& \frac{    4[8(D+3E)+5(f_{4,3}-f_{2,1}) ]\beta_{0}^{2}    +(f_{4,3}-f_{2,1})   [4\mathrm{ZFS}^{2} -(f_{4,3}-f_{2,1})^{2} -4  f_{3,2}^{2}      ]  }{96 \beta_{0}^{2}}\label{eq:eta}.
\end{eqnarray}
\end{widetext}
Thus, in general, if $D$ and $E$ values are known and three resonant energies are observable, the applied magnetic field strength can be extracted from Eq.(\ref{eq:B0}). In addition, if an uniaxial symmetry exists ($E \ll D$), because of $\eta \simeq D\cos^{2}{\theta}$, the polar angle can also be extracted from Eq.(\ref{eq:eta}).

\added{However, }when one wants to use these formulas to find the external magnetic field vector, one may face a problem: how can one determine which observed resonant energy is from which transition? This is explained in Fig.\ref{fig:TripletvsQuartet} and Fig.\ref{fig:transitionsatlowandhighfield}.
At a given magnetic field vector (e.g., \added{$g \mu$}$ B_{0} \gg \mathrm{ZFS} $), there will appear three strong transitions for $\mathrm{S=3/2}$ as in Fig.\ref{fig:transitionsatlowandhighfield}(a). The resonant energies are varying depending on the orientation for fixed $B_{0}$ and some of them even cross each other, thus it is hard to determine $f_{i,j}$ explicitly only taking into account these three transitions. However, in certain spin systems bound to certain localized defects in solids and at \added{a }certain magnetic field range, this ambiguity can be relaxed as will be shown in the following sections. We will focus on two distinguishable cases for \added{$g\mu_{B}$}$B_{0} \ll \mathrm{ZFS}$ and \added{$g\mu_{B}$}$B_{0} \gg \mathrm{ZFS}$. The case for \added{$g\mu_{B}$}$B_{0}\sim \mathrm{ZFS}$, however, will not be discussed because complex spectra appear due to \added{a }strong interaction among each eigenstate\deleted{s}~\cite{Degen2008} via e.g. the level anti-crossing as in the NV centers~\cite{He1993}, thus, high spin systems are not an appropriate sensor for this field range. In the following sections, we will consider the uniaxial symmetry for all the numerical simulations unless noticed. This will allow us to utilize Eq.(\ref{eq:eta}) to extract at least one orientation component, $\theta$.

\subsection{High field, \added{$g\mu_{B}$}$B_{0} \gg \mathrm{ZFS}$ \label{section:largefield}}
\begin{figure}
\includegraphics[width=1\columnwidth]{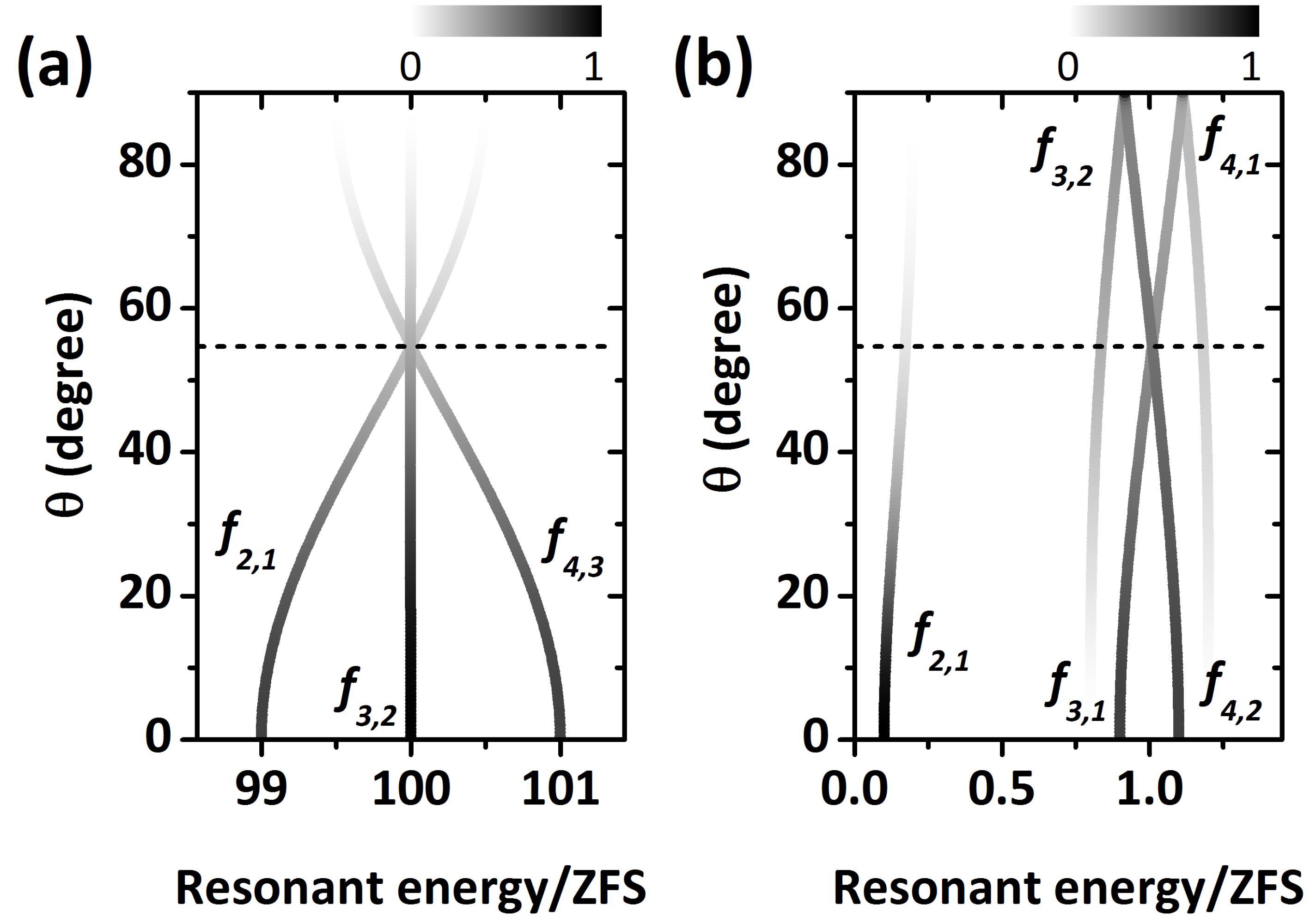}
\caption{\label{fig:transitionsatlowandhighfield}Orientation dependenc\replaced{e}{ies} of ESR transitions at (a) high \added{($g\mu_{B} B_{0}=100\times \mathrm{ZFS}$)} and (b) low \added{($g\mu_{B} B_{0}=\mathrm{ZFS}/10$)} magnetic field strength. The dashed lines indicate the magic angle $\theta_{m}$. The gray color scale depicts the normalized ESR transition probabilities for $\mathbf{B_{1}}\mathrm{ \parallel x-axis}$}.
\end{figure}
Figure\,\ref{fig:transitionsatlowandhighfield}(a) depicts how each transition evolves at varying orientation $\theta$ at high static magnetic field\added{s} (\added{$g\mu_{B}$}$B_{0}=100\times \mathrm{ZFS}$) together with ESR transition probabilities.
$\mathbf{B_{1}}$ is assumed to have only a\added{n} \replaced{x}{z}-axis component. At high field\added{s}, three transitions are visible. Without knowing the information about $B_{0}$, it is not possible to assign them correctly because as will be seen later, there appear\added{s} also three or even more than three resonances that originate from different transitions at low magnetic fields. However, one can guess $B_{0}$ roughly from the observed resonant energies because they are approximately proportional to $B_{0}$ at high field. For example, $f_{3,2}$ shows very weak orientation dependence thus can be used to estimate $B_{0}$. Once the $B_{0}$ scale is roughly guessed, one can try to assign the observed resonant energies. Because $f_{3,2}$ always stays in the middle of all resonance transitions, this can be explicitly determined. And because Eq.(\ref{eq:B0}) is invariant under switching of $f_{4,3}$ and $f_{2,1}$, $B_{0}$ can be unambiguously extracted. In contrast, Eq.(\ref{eq:eta}) will result in a systematic error if $f_{4,3}$ and $f_{2,1}$ are not correctly assigned. Thus\added{,} a magnetometer based on a $S=3/2$ system can be used only to extract $B_{0}$ at \added{a} high magnetic field. This is a disadvantage compared to $S=1$ because a similar equation to Eq.(\ref{eq:eta}) is also invariant under switching of two observed resonant energies~\cite{Balasubramanian2008}. This problem, however, may be overcome by manipulating the ZFS~\cite{Dolde2011}. If the electric dipole moment is large enough and their contribution to the ZFS Hamiltonian, the Stark-effect, is well-known, manipulation of the ZFS by either applying an electric field~\cite{Dolde2011} or pressure~\cite{Falk2014} along a favored direction can result in the shifts of each ESR transitions in different manner\added{s}. Thus monitoring the additional shift upon the change in ZFS, may allow to determine all the necessary transitions and subsequently Eq.(\ref{eq:eta}) can be used without systematic errors.


\subsection{Low field, \added{$g\mu_{B}$}$B_{0} \ll \mathrm{ZFS}$ \label{section:smallfield}}
Figure\,\ref{fig:transitionsatlowandhighfield}(b) shows that one can see up to five transitions at low field. At \added{a} small angle, there are three \deleted{most} dominant transitions, $f_{2,1}$, $f_{3,1}$, and $f_{4,2}$, and two additional transitions, $f_{3,2}$, and $f_{4,1}$, \added{which} arise at large angle. Alternate forms of Eq.(\ref{eq:B0}) and (\ref{eq:eta}) using only the most dominant transitions can be found for small angles (not shown). These forms, however, are not useful because $f_{4,2}$ and $f_{3,1}$ are changing their relative positions at \added{a} larger angle. Instead, $f_{4,1}$ and $f_{3,2}$ can be used since their relative positions are not changing and in certain $S=3/2$ systems, e.g. $\mathrm{V_{Si}}$ in SiC (see section \ref{section:siliconvacancy}), these transitions show good intensities at every orientation~\cite{Simin2015}. The other useful formulas can be found by plugging $f_{4,3}=f_{4,1}-f_{3,2}-f_{2,1}$ into Eqs.(\ref{eq:B0}) and (\ref{eq:eta}) as
\begin{widetext}
\begin{eqnarray}
&\beta_{0}^{2} &= \frac{(\sqrt{3} f_{avg} + f_{2,1})^2-2f_{4,1} f_{3,2}+(1-\sqrt{3}) f_{3,2} f_{2,1} -(1+\sqrt{3}) f_{41} f_{21}-\mathrm{ZFS}^{2}      }{5}\added{,}\label{eq:B0forsmallB0without4231}\\
&\eta &= \frac{1}{96 \beta_{0}^{2}}
\{ 4[8(D+3E)+5\Delta f_{out}-10f_{2,1}]\beta_{0}^{2}
-8 f_{2,1} (\mathrm{ZFS}^{2}-f_{2,1}^{2}-f_{3,2}^{2})
\nonumber \\
&&+\Delta f_{out} [ 4 ({\mathrm{ZFS}}^{2} -3 f_{2,1}^{2}-f_{3,2}^{2}) +\Delta f_{out} (6 f_{2,1}-\Delta f_{out})                              ] \}\added{,}
\label{eq:etaforsmallB0without4231}
\end{eqnarray}
\end{widetext}
where $\Delta f_{out}\equiv f_{4,1}-f_{3,2}$ and $f_{avg}\equiv (f_{3,1}+f_{4,2})/2=(f_{3,2}+f_{4,1})/2$. These are useful, because if $f_{4,1}$ and $f_{3,2}$ are observable together with $f_{4,2}$ and $f_{3,1}$, one always can unambiguously determine the two outermost transitions, and $f_{4,2}$ and $f_{3,1}$ are not in use or necessary only for calculating $f_{avg}$.

So far, the strategies to use $S=3/2$ systems as a DC vector magnetometer have been discussed in both high and low magnetic field ranges. Though only $B_{0}$ can be obtained at high field\added{s} using only a conventional experimental method, both $B_{0}$ and polar angle can be determined at low field. However, in many high spin systems bound to localized defects in solid\added{s}, spin-dependent intersystem-crossing may induce a strong polarization into specific spin states as in $\mathrm{V_{Si}}$ in SiC~\cite{Mizuochi2002,Isoya2008}. Thus, some transitions may be hardly observable. In addition, because only the polar angle can be obtained, it is still not possible to realize a genuine vector magnetometry. In the following sections, we will present $S=3/2$ spins of $\mathrm{V_{Si}}$ in SiC as a model system and discuss the practical usage of them and a possible way to use them as a vector magnetometer.

\section{Silicon vacancy spins in silicon carbide as a DC vector magnetome\deleted{n}ter\label{section:siliconvacancy}}
We present $\mathrm{V_{Si}}$ in SiC as a model system to provide explanations about how the formulas, found in previous sections, can be used to experimentally reconstruct the applied external magnetic field vector. Because spin properties are different depending on the polytype, here we discuss only a specific polytype, namely 4H-SiC. In addition, because there exist two inequivalent lattice sites, there appear two different silicon vacancies with different ZFS, and we choose only one of them known as $T_{V2a}$ center~\cite{Sorman2000,Janzen2009}. In the $T_{V2a}$ center, it is known that there exist an uniaxial symmetry around the c-axis thus $E\ll D$ and $\mathrm{ZFS}\added{/h}\simeq 2D\added{/h}\simeq 70~\mathrm{MHz}$~\cite{Sorman2000,Janzen2009}. It is also known that optical polarization results in equal populations in two substates, $|m_{S}=\pm 1/2 \rangle$.
\deleted{Because this polarization is almost independent of the static magnetic field orientation, resonant transitions between these two states have not been observed in the orientation dependence of continuous wave (cw) experiment at cryogenic~\cite{Sorman2000,Soltamov2012,Mizuochi2002} and room temperature~\cite{Simin2015,Kraus2014}.}
\added{This is responsible for the absence of a transition between $|m_{S}=+1/2 \rangle$ and $|m_{S}= -1/2 \rangle$ while aother two transitions, between $|m_{S}= +3/2 \rangle$ and $|m_{S}= +1/2 \rangle$, and between $|m_{S}= -3/2 \rangle$ and $|m_{S}= -1/2 \rangle$ are observable in the ESR spectra of $T_{V2a}$ for $\mathbf{B_{0}}||$c-axis~\cite{Widmann2015, Soltamov2012, Sorman2000, Mizuochi2002, Kraus2014a, Kraus2014, Simin2015, Baranov2011}. In the $\mathbf{B_{0}}$ orientation dependence at low ~\cite{Simin2015} and high magnetic fields~\cite{Sorman2000, Mizuochi2002, Soltamov2012, Kraus2014a}, one of the allowed transitions, corresponding to the transition between $|m_{S}=+1/2 \rangle$ and $|m_{S}= -1/2 \rangle$ for $\mathbf{B_{0}}||$c-axis,  \replaced{has}{have} not been observed probably due to that this equal population is somehow maintained.}
This will prevent Eq.(\ref{eq:B0}), (\ref{eq:eta}), (\ref{eq:B0forsmallB0without4231}), and (\ref{eq:etaforsmallB0without4231}) from being used because $f_{3,2}$ at high field\added{s} and $f_{2,1}$ at low field\added{s} will not be observable. This transition, however, can become visible once electron-electron double resonance (ELDOR) is applied. The population difference between $|m_{S}=\pm 1/2 \rangle$ states can be induced by applying e.g. a resonant $\pi$ pulse between $|m_{S}=+3/2 \rangle$ and $|m_{S}=+1/2 \rangle$ (or $|m_{S}=-1/2 \rangle$ and $|m_{S}=-3/2 \rangle$) states which enable detection of this missing transition~\cite{Isoya2008}. This will allow unambiguous determination of one transition $f_{2,1}$ at low field\added{s} or $f_{3,2}$ at high field\added{s} experimentally. 

At high magnetic field (e.g., $B_{0}\sim 300$\replaced{$~mT$}{$0\,G$}) as in Fig.\ref{fig:transitionsatlowandhighfield}(a), two outer transitions, $f_{4,3}$ and $f_{2,1}$ have been observed experimentally at almost all orientation\added{s} at both cryogenic ~\cite{Mizuochi2002,Sorman2000} and room temperature~\cite{Kraus2014a} except the \replaced{central}{center} peak. The \replaced{central}{center} peak is observable by ELDOR experiments~\cite{Isoya2008}, thus Eq.(\ref{eq:B0}) can be used as explained in section (\ref{section:largefield}). However, the polar angle, $\theta$, can be determined from Eq.(\ref{eq:eta}) only at small angles because of the ambiguity on determining $f_{4,3}$ and $f_{2,1}$ at larger angle\added{s}.

ESR spectra of $T_{V2a}$ centers at \added{a} low magnetic field (e.g. sub-mT) as in Fig.\ref{fig:transitionsatlowandhighfield}(b) allow an unambiguous determination of both $B_{0}$ and polar angle as long as the ELDOR can be used to determine $f_{2,1}$ as explained in section (\ref{section:smallfield}). However, one can consider another case in which either ELDOR experiments are not available or $f_{2,1}$ is hardly observed in the ELDOR spectrum. In such a case, if $f_{4,1}$ and $f_{3,2}$ are observable,
using relations $f_{3,1}+f_{4,2}=f_{3,2}+f_{4,1}$, $f_{4,3}=f_{4,2}-f_{3,2}$, and $f_{2,1}=f_{3,1}-f_{3,2}$ from Eq.(\ref{eq:quarteteigenvalues}), we again obtain alternative forms of Eqs.(\ref{eq:B0}) and (\ref{eq:eta}) as
\begin{widetext}
\begin{eqnarray}
&&B_{0}^{2} = \frac{(\sqrt{3} f_{avg} + f_{3,2})^2-f_{4,2} f_{3,1}-2(\sqrt{3}+1)f_{3,2}f_{avg} -\mathrm{ZFS}^{2}      }{5}\added{,}\label{eq:B0forsmallB0without21using32}\\
&&\eta = \frac
{
[32(D+3E) +20 \Delta f_{in}   ] B_{0}^{2}
+\Delta f_{in} ( 4 {\mathrm{ZFS}}^{2} -\Delta f_{i}^{2} - 4 f_{3,2}^{2}       )
 }
{96 B_{0}^{2}}\added{,}\label{eq:etaforsmallB0without21using32}
\end{eqnarray}
\end{widetext}
where $\Delta f_{in}\equiv f_{4,2}-f_{3,1}$. Note that $f_{3,2}$ appears in both formula\added{s} but because it is always the lowest energy transition, this can be explicitly determined. Similarly, one can find additional alternatives using $f_{4,1}$ instead of $f_{3,2}$.
Therefore, even if ELDOR is not available, as long as either $f_{4,1}$ or $f_{3,2}$ is observable together with $f_{4,2}$ and $f_{3,1}$, $B_{0}$ can be extracted using Eq.(\ref{eq:B0forsmallB0without21using32}) because it is invariant under switching $f_{4,2}$ and $f_{3,1}$. This scheme is feasible since $f_{4,1}$ and $f_{3,2}$ are observable from $\mathrm{T_{V2a}}$ in SiC by cw method\added{s} with a decent signal strength at sub-mT as recently reported~\cite{Simin2015}. Eq.(\ref{eq:etaforsmallB0without21using32}), however, still cannot provide an unambiguous way to determine the polar angle because of $\Delta f_{in}$ which changes signs if $f_{4,2}$ and $f_{3,1}$ are not correctly determined.

So far, the strategies to use $\mathrm{V_{Si}}$ in SiC as a vector magnetometer has been discussed. While the magnetic field strength can be extracted in both high and low magnetic field range, the orientation can be extracted only if there exists an uniaxial symmetry at \added{a} low magnetic field, and the azimuthal angle cannot be determined in any case. Note that the S=1 system with the uniaxial symmetry also can provide only the polar angle. But in the case of the NV center in diamond, because the NV centers can be in \replaced{four}{three} different orientations along the diamond bond axes, one can determine both the polar and azimuthal angles from the shift of transitions \replaced{of}{from} the inequivalently oriented NV centers. Similarly, $\mathrm{V_{Si}}$ in inequivalent lattice sites, e.g. $\mathrm{T_{V1a}}$ and $\mathrm{T_{V2a}}$ in 4H-SiC, and $\mathrm{T_{V1a}}$, $\mathrm{T_{V2a}}$ and $\mathrm{T_{V3a}}$ in 6H-SiC~\cite{Janzen2009} can also be utilized. However\added{,} ESR spectra of $\mathrm{T_{V1a}}$ and $\mathrm{T_{V3a}}$ are hardly visible at room temperature~\cite{Janzen2009,Baranov2011,Soltamov2012}. Thus, an alternate method relying only on the $\mathrm{T_{V2a}}$ center that can be used for any magnetic field strength at room temperature is necessary. In the next section, another method using a magic angle that allows for the use of S=3/2 as a vector magnetometer will be discussed.

\section{Vector magnetometry using magic angle \label{section:magic_angle}}
We start from the eigenvalue equation in Eq.(\ref{eq:eigenvalue_equation}). In this equation, one can find terms including  $\mathrm{(3 cos^{2}{\theta} -1)}$, which becomes zero at the magic angle $\theta_{m}\simeq 54.7\,\degree$. Eq.(\ref{eq:eigenvalue_equation}) can be simplified for $\theta = \theta_{m} $ and $E\ll D$ as,
\begin{equation}\label{eq:eigenvealueformula_magic_angle}
\lambda^{4}-(2D^{2}+\frac{5}{2}\beta_{0}^{2})\lambda^{2}+\frac{9}{16}\beta_{0}^{4}+D^{4} -\frac{1}{2}D^{2}\beta_{0}^{2}=0\added{,}
\end{equation}
and the eigenvalues are simply
\begin{equation}\label{eq:eigenvealue_magic_angle}
\lambda=\pm \frac{1}{2} \sqrt{4 D^{2} + 5 \beta_{0}^{2} \pm 4 \sqrt{ 3 \beta_{0}^{2} D_{0}^{2}   +\beta_{0}^{4}     }       }\added{,}
\end{equation}
as depicted in Fig.\ref{fig:eigenvaluesvsB0}(c). For high $B_{0}$, these can be again approximated as
\begin{equation}\label{eq:eigenvealue_magic_angle_highfield}
\lambda=\pm \frac{3}{2}\beta_{0}\,\,\, \mathrm{or}\,\,\, \pm \frac{1}{2}\beta_{0} \,\,\,(\mathrm{for}\,\,\,
g \mu_{B} B_{0} \gg \mathrm{ZFS}).
\end{equation}
Thus, at $\theta_{m}$, we obtain $|\lambda_{1}|=|\lambda_{4}|$ and $|\lambda_{2}|=|\lambda_{3}|$, and can see the least number of transitions as \deleted{can be }seen in Fig.\ref{fig:transitionsatlowandhighfield}. We can use this aspect to use $\mathrm{S=3/2}$ system as a vector magnetometer. If the spin sensor is being rotated around an axis and the orientation between the c-axis and the rotational axis is fixed to $\theta_{m}$, one expects to see the least number of transitions whose widths are the narrowest when the rotational axis is aligned to the applied external magnetic field. In contrast, when the rotational axis is misaligned, very broad ESR transitions appear due to orientation sweeping.
\begin{figure}
\includegraphics[width=1\columnwidth]{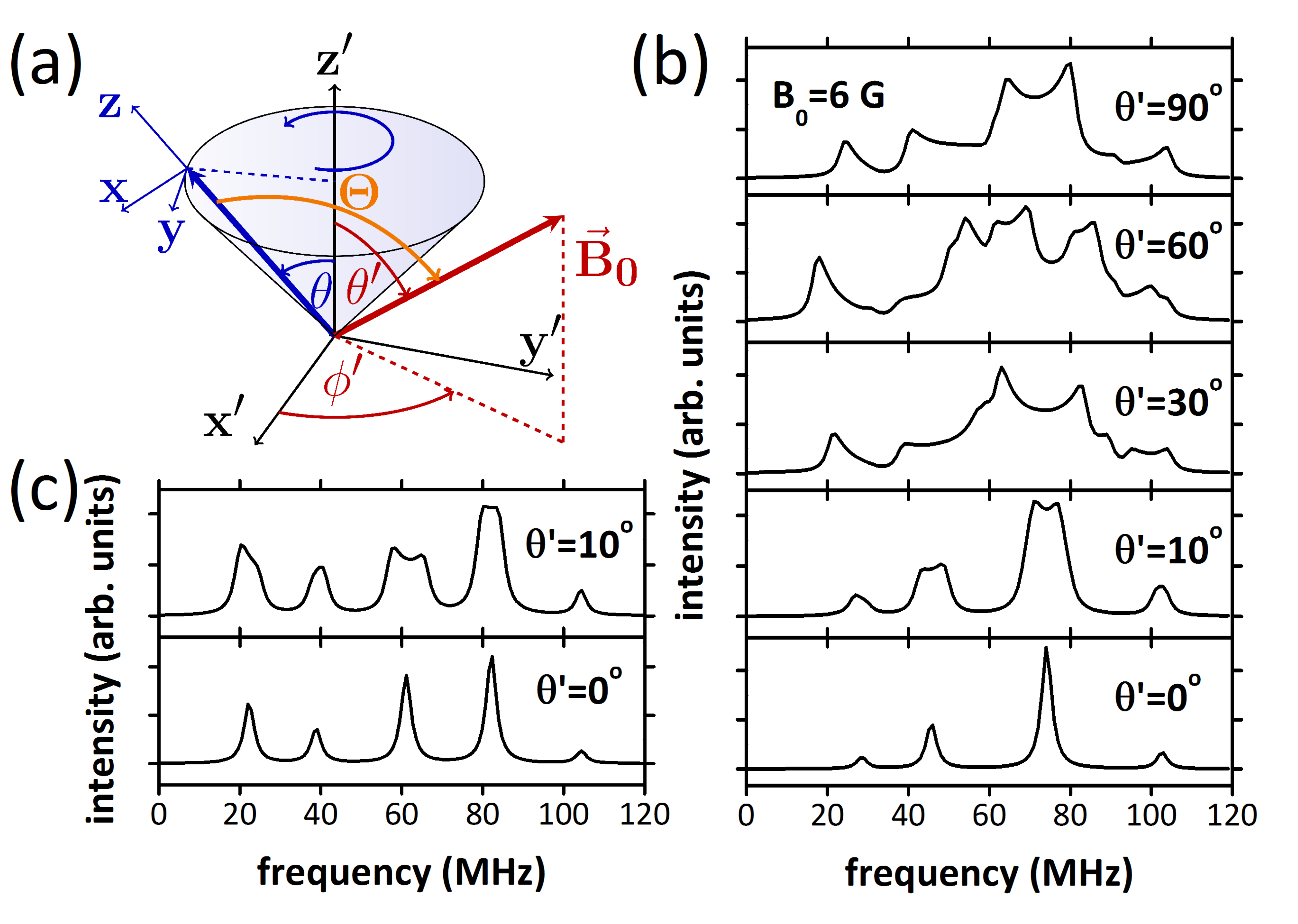}
\caption{\label{fig:magicangle}(Color online) Magic angle DC magnetometry based on a S=3/2 spin system. (a) describes the frames used for the numerical simulation. See text for details. (b) shows the numerical simulations of the cw ESR spectra when the S=3/2 spin is \replaced{misaligned}{misoriented} by $\theta=\theta_{m}$ relative to the rotational axis at various orientations of $\mathbf{B_{0}}$ vector in $\mathrm{y^{\prime}-z^{\prime}}$ plane at low field $B_{0}\mathrm{=6\,G}$. For the simulation, the ZFS of $T_{V2a}$ center in SiC, $\mathrm{ZFS}\added{/h}\simeq2D\added{/h}\simeq70\,\mathrm{MHz}$ is assumed. (c) shows the same simulation but for \added{the} case when $\theta \neq \theta_{m}$ \replaced{and}{but} $\theta=30\,\degree$.}
\end{figure}

Fig\replaced{ure }{.}\ref{fig:magicangle} describes such an experiment in which the SiC crystal is attached to a rotational axis forming $\theta_{m}$ with respect to the c-axis of the crystal. The resonant RF can be applied using a miniature coil surrounding either the rotational axis or the SiC crystal,\added{ similar to what has been suggested for the quantum gyroscope based on the NV center~\cite{Ajoy2012}}. For the detection, a small sized ESR cavity can be used for conventional ESR detection. Fiber coupling also can be considered for ODMR. If an electrically detected magnetic resonance is possible, which has recently been shown in high spin systems~\cite{2015arXiv150207551B} and also in SiC~\cite{Cochrane2012}, additional small circuits can be utilized. $\mathrm{B_{1}}$ field modulation can be used to enhance the signal to noise ratio~\cite{Lee2012} because the ELDOR experiment which requires $\mathrm{B_{1}}$ pulses is not necessary \replaced{for}{in} this experiment. For the simulation of ESR spectra in this experiment, a laboratory frame is assumed: the $\mathrm{z^{\prime}-axis}$ is set to the rotational axis. The angle ($\Theta$) between the c-axis of the crystal, rotating with a constant speed around the $\mathrm{z^{\prime}-axis}$, and the external magnetic field can be derived and replace $\theta$ in Eq.(\ref{eq:eigenvalue_equation}). For convenience, the RF field is assumed to be in the x-axis of the rotating frame. By assuming a Lorent\replaced{zi}{iz}an lineshape with 3 MHz FWHM, the numerically simulated ESR spectra at varying $\theta^{\prime}$ while $\phi^{\prime}$ is fixed to $90\,\degree$ are simulated for \added{a} low field ($B_{0}=6~G$) as shown in Fig.\ref{fig:magicangle}(b). As expected, when $\theta^{\prime}=0$ equivalently $\Theta=\theta_{m}$ or $\bf{B_{0}}\parallel \mathrm{z^{\prime}-axis}$, the narrowes\added{t} transitions are found while seriously broadened peaks like a powder pattern appear\added{s} when misaligned ($\theta^{\prime}\neq0$). Note that in order to present a general $\mathrm{S=3/2}$ case, $f_{2,1}$ is assumed to be visible in cw ESR spectra at low field. Therefore by monitoring the linewidth of the observed transition spectra while moving the rotational axis, $z^{\prime}$-axis, one can explicitly find the orientation of the external magnetic field. The field strength also can be extracted from the observed resonant energy of the strongest transition using Eq.(\ref{eq:eigenvealue_magic_angle}). For very small field \added{$g\mu_{B}$}$B_{0}\ll \mathrm{ZFS}$, $f_{3,1}=f_{4,2} \approx$
\added{$g\mu_{B}$}$B_{0}+4$\added{$g\mu_{B}$}$B_{0}D^{2}/3$. We also can observe spectra consisting of the narrow transitions even if $\theta\neq\theta_{m}$ as long as $\theta^{\prime}=0$. However, because many transitions whose intensities are comparable to each other appear as in Fig.\ref{fig:magicangle}(c), it is more convenient to use the magic angle because the most dominant transition is easily distinguishable.

\section{Conclusion \label{section:conclusion}}
We have shown that using the $\mathrm{V_{Si}}$ in SiC as a model system, S=3/2 electronic spins with the uniaxial symmetry can be used to find the strength and polar angle of the applied external magnetic field if at least three ESR transitions can be found experimentally and the ZFS parameters are known.
At \added{a} high $B_{0}$ field (\added{$g\mu_{B}$}$B_{0}\gg \mathrm{ZFS}$), 
$B_{0}$ can be obtained from the observed ESR spectra but the polar angle cannot be determined due to the ambiguity of differentiating two outer transitions.
In contrast, at low \added{$g\mu_{B}$}$B_{0}$ ($\ll \mathrm{ZFS}$), as long as one can explicitly identify at least three transitions including the allowed lowest energy transition, the external magnetic field vector can be reconstructed. In the field strength comparable to the ZFS, it is hard to find a\deleted{n} useful scheme because very complex patterns appear due to mixing of some of the eigenstates. In the case of the NV centers in diamond (ZFS\added{/h}=2.87 GH\replaced{z}{Z}), this missing range is around \replaced{$\sim100 \, mT$}{$\sim1000 \, G$}. The $V_{Si}$ in SiC can fill out this gap since its ZFS is quite small ($\mathrm{ZFS}\added{/h}\sim 100 \,\mathrm{MHz}$) thus this magnetic field range can be considered as a high field range in which the three necessary transitions are well observable~\cite{Isoya2008,Kraus2014a}, and at least the field strength can be experimentally determined. When the $V_{Si}$ in SiC is used to realize such schemes at sub-mT, if the lowest transition energy is observable by ELDOR, one can determine both $B_{0}$ and $\theta$ without ambiguity. Even if ELDOR is not available, thanks to the additional transitions that appear at low field\added{s}, the field strength can be determined.

The magic angle terms in the eigenvalue equation allow for an alternative method to use S=3/2 systems as a DC vector magnetometer. If the S=3/2 spins fixed in a crystal can be rotated around the rotational axis, the unambiguous determination of the applied magnetic field vector is feasible by monitoring the linewidth of the observed ESR spectra while the symmetry axis of the crystal is oriented at $\theta_{m}$ relative to the rotational axis and the rotational axis is moving. This configuration also can be realized by producing an array of the crystals such that the symmetry axes of each crystal form a cone whose opening angle is twice the magic angle.

These findings provide a better understanding of the S=3/2 electronic spin Hamiltonian\added{,} especially at low field\added{s}. They also provide an outlook for the application of $\mathrm{V_{Si}}$ in SiC to quantum magnetometry which is promising thanks to the electrical properties of SiC, which outstand the host material of the NV centers, and the mature fabrication technology, which allows an efficient fabrication of electronic devices even at the atomic scale~\cite{Lohrmann2015}.

\begin{acknowledgments}
We thank Torsten Rendler, Seoyoung Paik, Thomas Wolf, Matthias Widmann for helpful discussions, and Nathan Chejanovsky for helping to prepare this manuscript. We acknowledge funding by the DFG via priority programme 1601 and the EU via ERC grant SQUTEC and Diadems as well as the Max Planck Society.
\end{acknowledgments}



\begin{thebibliography}{48}%
\makeatletter
\providecommand \@ifxundefined [1]{%
 \@ifx{#1\undefined}
}%
\providecommand \@ifnum [1]{%
 \ifnum #1\expandafter \@firstoftwo
 \else \expandafter \@secondoftwo
 \fi
}%
\providecommand \@ifx [1]{%
 \ifx #1\expandafter \@firstoftwo
 \else \expandafter \@secondoftwo
 \fi
}%
\providecommand \natexlab [1]{#1}%
\providecommand \enquote  [1]{``#1''}%
\providecommand \bibnamefont  [1]{#1}%
\providecommand \bibfnamefont [1]{#1}%
\providecommand \citenamefont [1]{#1}%
\providecommand \href@noop [0]{\@secondoftwo}%
\providecommand \href [0]{\begingroup \@sanitize@url \@href}%
\providecommand \@href[1]{\@@startlink{#1}\@@href}%
\providecommand \@@href[1]{\endgroup#1\@@endlink}%
\providecommand \@sanitize@url [0]{\catcode `\\12\catcode `\$12\catcode
  `\&12\catcode `\#12\catcode `\^12\catcode `\_12\catcode `\%12\relax}%
\providecommand \@@startlink[1]{}%
\providecommand \@@endlink[0]{}%
\providecommand \url  [0]{\begingroup\@sanitize@url \@url }%
\providecommand \@url [1]{\endgroup\@href {#1}{\urlprefix }}%
\providecommand \urlprefix  [0]{URL }%
\providecommand \Eprint [0]{\href }%
\providecommand \doibase [0]{http://dx.doi.org/}%
\providecommand \selectlanguage [0]{\@gobble}%
\providecommand \bibinfo  [0]{\@secondoftwo}%
\providecommand \bibfield  [0]{\@secondoftwo}%
\providecommand \translation [1]{[#1]}%
\providecommand \BibitemOpen [0]{}%
\providecommand \bibitemStop [0]{}%
\providecommand \bibitemNoStop [0]{.\EOS\space}%
\providecommand \EOS [0]{\spacefactor3000\relax}%
\providecommand \BibitemShut  [1]{\csname bibitem#1\endcsname}%
\let\auto@bib@innerbib\@empty
\bibitem [{\citenamefont {Doherty}\ \emph {et~al.}(2013)\citenamefont
  {Doherty}, \citenamefont {Manson}, \citenamefont {Delaney}, \citenamefont
  {Jelezko}, \citenamefont {Wrachtrup},\ and\ \citenamefont
  {Hollenberg}}]{Doherty2013}%
  \BibitemOpen
  \bibfield  {author} {\bibinfo {author} {\bibfnamefont {M.~W.}\ \bibnamefont
  {Doherty}}, \bibinfo {author} {\bibfnamefont {N.~B.}\ \bibnamefont {Manson}},
  \bibinfo {author} {\bibfnamefont {P.}~\bibnamefont {Delaney}}, \bibinfo
  {author} {\bibfnamefont {F.}~\bibnamefont {Jelezko}}, \bibinfo {author}
  {\bibfnamefont {J.}~\bibnamefont {Wrachtrup}}, \ and\ \bibinfo {author}
  {\bibfnamefont {L.~C.~L.}\ \bibnamefont {Hollenberg}},\ }\href {\doibase
  http://dx.doi.org/10.1016/j.physrep.2013.02.001} {\bibfield  {journal}
  {\bibinfo  {journal} {Physics Reports}\ }\textbf {\bibinfo {volume} {528}},\
  \bibinfo {pages} {1} (\bibinfo {year} {2013})}\BibitemShut {NoStop}%
\bibitem [{\citenamefont {Schirhagl}\ \emph {et~al.}(2014)\citenamefont
  {Schirhagl}, \citenamefont {Chang}, \citenamefont {Loretz},\ and\
  \citenamefont {Degen}}]{Schirhagl2014}%
  \BibitemOpen
  \bibfield  {author} {\bibinfo {author} {\bibfnamefont {R.}~\bibnamefont
  {Schirhagl}}, \bibinfo {author} {\bibfnamefont {K.}~\bibnamefont {Chang}},
  \bibinfo {author} {\bibfnamefont {M.}~\bibnamefont {Loretz}}, \ and\ \bibinfo
  {author} {\bibfnamefont {C.~L.}\ \bibnamefont {Degen}},\ }\href {\doibase
  10.1146/annurev-physchem-040513-103659} {\bibfield  {journal} {\bibinfo
  {journal} {Annual Review of Physical Chemistry}\ }\textbf {\bibinfo {volume}
  {65}},\ \bibinfo {pages} {83} (\bibinfo {year} {2014})}\BibitemShut {NoStop}%
\bibitem [{\citenamefont {S\"{o}rman}\ \emph {et~al.}(2000)\citenamefont
  {S\"{o}rman}, \citenamefont {Son}, \citenamefont {Chen}, \citenamefont
  {Kordina}, \citenamefont {Hallin},\ and\ \citenamefont
  {Janz\'{e}n}}]{Sorman2000}%
  \BibitemOpen
  \bibfield  {author} {\bibinfo {author} {\bibfnamefont {E.}~\bibnamefont
  {S\"{o}rman}}, \bibinfo {author} {\bibfnamefont {N.~T.}\ \bibnamefont {Son}},
  \bibinfo {author} {\bibfnamefont {W.~M.}\ \bibnamefont {Chen}}, \bibinfo
  {author} {\bibfnamefont {O.}~\bibnamefont {Kordina}}, \bibinfo {author}
  {\bibfnamefont {C.}~\bibnamefont {Hallin}}, \ and\ \bibinfo {author}
  {\bibfnamefont {E.}~\bibnamefont {Janz\'{e}n}},\ }\href
  {http://link.aps.org/doi/10.1103/PhysRevB.61.2613} {\bibfield  {journal}
  {\bibinfo  {journal} {Physical Review B}\ }\textbf {\bibinfo {volume} {61}},\
  \bibinfo {pages} {2613} (\bibinfo {year} {2000})}\BibitemShut {NoStop}%
\bibitem [{\citenamefont {Janz\'{e}n}\ \emph {et~al.}(2009)\citenamefont
  {Janz\'{e}n}, \citenamefont {Gali}, \citenamefont {Carlsson}, \citenamefont
  {G\"{a}llstr\"{o}m}, \citenamefont {Magnusson},\ and\ \citenamefont
  {Son}}]{Janzen2009}%
  \BibitemOpen
  \bibfield  {author} {\bibinfo {author} {\bibfnamefont {E.}~\bibnamefont
  {Janz\'{e}n}}, \bibinfo {author} {\bibfnamefont {A.}~\bibnamefont {Gali}},
  \bibinfo {author} {\bibfnamefont {P.}~\bibnamefont {Carlsson}}, \bibinfo
  {author} {\bibfnamefont {A.}~\bibnamefont {G\"{a}llstr\"{o}m}}, \bibinfo
  {author} {\bibfnamefont {B.}~\bibnamefont {Magnusson}}, \ and\ \bibinfo
  {author} {\bibfnamefont {N.~T.}\ \bibnamefont {Son}},\ }\href {\doibase
  http://dx.doi.org/10.1016/j.physb.2009.09.023} {\bibfield  {journal}
  {\bibinfo  {journal} {Physica B: Condensed Matter}\ }\textbf {\bibinfo
  {volume} {404}},\ \bibinfo {pages} {4354} (\bibinfo {year}
  {2009})}\BibitemShut {NoStop}%
\bibitem [{\citenamefont {Widmann}\ \emph {et~al.}(2015)\citenamefont
  {Widmann}, \citenamefont {Lee}, \citenamefont {Rendler}, \citenamefont {Son},
  \citenamefont {Fedder}, \citenamefont {Paik}, \citenamefont {Yang},
  \citenamefont {Zhao}, \citenamefont {Yang}, \citenamefont {Booker},
  \citenamefont {Denisenko}, \citenamefont {Jamali}, \citenamefont
  {Momenzadeh}, \citenamefont {Gerhardt}, \citenamefont {Ohshima},
  \citenamefont {Gali}, \citenamefont {Janz\'{e}n},\ and\ \citenamefont
  {Wrachtrup}}]{Widmann2015}%
  \BibitemOpen
  \bibfield  {author} {\bibinfo {author} {\bibfnamefont {M.}~\bibnamefont
  {Widmann}}, \bibinfo {author} {\bibfnamefont {S.-Y.}\ \bibnamefont {Lee}},
  \bibinfo {author} {\bibfnamefont {T.}~\bibnamefont {Rendler}}, \bibinfo
  {author} {\bibfnamefont {N.~T.}\ \bibnamefont {Son}}, \bibinfo {author}
  {\bibfnamefont {H.}~\bibnamefont {Fedder}}, \bibinfo {author} {\bibfnamefont
  {S.}~\bibnamefont {Paik}}, \bibinfo {author} {\bibfnamefont {L.-P.}\
  \bibnamefont {Yang}}, \bibinfo {author} {\bibfnamefont {N.}~\bibnamefont
  {Zhao}}, \bibinfo {author} {\bibfnamefont {S.}~\bibnamefont {Yang}}, \bibinfo
  {author} {\bibfnamefont {I.}~\bibnamefont {Booker}}, \bibinfo {author}
  {\bibfnamefont {A.}~\bibnamefont {Denisenko}}, \bibinfo {author}
  {\bibfnamefont {M.}~\bibnamefont {Jamali}}, \bibinfo {author} {\bibfnamefont
  {S.~A.}\ \bibnamefont {Momenzadeh}}, \bibinfo {author} {\bibfnamefont
  {I.}~\bibnamefont {Gerhardt}}, \bibinfo {author} {\bibfnamefont
  {T.}~\bibnamefont {Ohshima}}, \bibinfo {author} {\bibfnamefont
  {A.}~\bibnamefont {Gali}}, \bibinfo {author} {\bibfnamefont {E.}~\bibnamefont
  {Janz\'{e}n}}, \ and\ \bibinfo {author} {\bibfnamefont {J.}~\bibnamefont
  {Wrachtrup}},\ }\href {\doibase 10.1038/nmat4145} {\bibfield  {journal}
  {\bibinfo  {journal} {Nat Mater}\ }\textbf {\bibinfo {volume} {14}},\
  \bibinfo {pages} {164} (\bibinfo {year} {2015})},\ \Eprint
  {http://arxiv.org/abs/1407.0180} {arXiv:1407.0180} \BibitemShut {NoStop}%
\bibitem [{\citenamefont {Christle}\ \emph {et~al.}(2015)\citenamefont
  {Christle}, \citenamefont {Falk}, \citenamefont {Andrich}, \citenamefont
  {Klimov}, \citenamefont {Hassan}, \citenamefont {Son}, \citenamefont
  {Janz\'{e}n}, \citenamefont {Ohshima},\ and\ \citenamefont
  {Awschalom}}]{Christle2015}%
  \BibitemOpen
  \bibfield  {author} {\bibinfo {author} {\bibfnamefont {D.~J.}\ \bibnamefont
  {Christle}}, \bibinfo {author} {\bibfnamefont {A.~L.}\ \bibnamefont {Falk}},
  \bibinfo {author} {\bibfnamefont {P.}~\bibnamefont {Andrich}}, \bibinfo
  {author} {\bibfnamefont {P.~V.}\ \bibnamefont {Klimov}}, \bibinfo {author}
  {\bibfnamefont {J.~U.}\ \bibnamefont {Hassan}}, \bibinfo {author}
  {\bibfnamefont {N.~T.}\ \bibnamefont {Son}}, \bibinfo {author} {\bibfnamefont
  {E.}~\bibnamefont {Janz\'{e}n}}, \bibinfo {author} {\bibfnamefont
  {T.}~\bibnamefont {Ohshima}}, \ and\ \bibinfo {author} {\bibfnamefont
  {D.~D.}\ \bibnamefont {Awschalom}},\ }\href {\doibase 10.1038/NMAT4144}
  {\bibfield  {journal} {\bibinfo  {journal} {Nat Mater}\ }\textbf {\bibinfo
  {volume} {14}},\ \bibinfo {pages} {160} (\bibinfo {year} {2015})},\ \Eprint
  {http://arxiv.org/abs/1406.7325} {arXiv:1406.7325} \BibitemShut {NoStop}%
\bibitem [{\citenamefont {Weber}\ \emph {et~al.}(2010)\citenamefont {Weber},
  \citenamefont {Koehl}, \citenamefont {Varley}, \citenamefont {Janotti},
  \citenamefont {Buckley}, \citenamefont {{Van de Walle}},\ and\ \citenamefont
  {Awschalom}}]{Weber2010a}%
  \BibitemOpen
  \bibfield  {author} {\bibinfo {author} {\bibfnamefont {J.~R.}\ \bibnamefont
  {Weber}}, \bibinfo {author} {\bibfnamefont {W.~F.}\ \bibnamefont {Koehl}},
  \bibinfo {author} {\bibfnamefont {J.~B.}\ \bibnamefont {Varley}}, \bibinfo
  {author} {\bibfnamefont {A.}~\bibnamefont {Janotti}}, \bibinfo {author}
  {\bibfnamefont {B.~B.}\ \bibnamefont {Buckley}}, \bibinfo {author}
  {\bibfnamefont {C.~G.}\ \bibnamefont {{Van de Walle}}}, \ and\ \bibinfo
  {author} {\bibfnamefont {D.~D.}\ \bibnamefont {Awschalom}},\ }\href {\doibase
  10.1073/pnas.1003052107} {\bibfield  {journal} {\bibinfo  {journal}
  {Proceedings of the National Academy of Sciences}\ }\textbf {\bibinfo
  {volume} {107}},\ \bibinfo {pages} {8513} (\bibinfo {year}
  {2010})}\BibitemShut {NoStop}%
\bibitem [{\citenamefont {Koehl}\ \emph {et~al.}(2011)\citenamefont {Koehl},
  \citenamefont {Buckley}, \citenamefont {Heremans}, \citenamefont {Calusine},\
  and\ \citenamefont {Awschalom}}]{Koehl2011a}%
  \BibitemOpen
  \bibfield  {author} {\bibinfo {author} {\bibfnamefont {W.~F.}\ \bibnamefont
  {Koehl}}, \bibinfo {author} {\bibfnamefont {B.~B.}\ \bibnamefont {Buckley}},
  \bibinfo {author} {\bibfnamefont {F.~J.}\ \bibnamefont {Heremans}}, \bibinfo
  {author} {\bibfnamefont {G.}~\bibnamefont {Calusine}}, \ and\ \bibinfo
  {author} {\bibfnamefont {D.~D.}\ \bibnamefont {Awschalom}},\ }\href {\doibase
  http://www.nature.com/nature/journal/v479/n7371/abs/nature10562.html\#supplementary-information}
  {\bibfield  {journal} {\bibinfo  {journal} {Nature}\ }\textbf {\bibinfo
  {volume} {479}},\ \bibinfo {pages} {84} (\bibinfo {year} {2011})}\BibitemShut
  {NoStop}%
\bibitem [{\citenamefont {Baranov}\ \emph {et~al.}(2011)\citenamefont
  {Baranov}, \citenamefont {Bundakova}, \citenamefont {Soltamova},
  \citenamefont {Orlinskii}, \citenamefont {Borovykh}, \citenamefont
  {Zondervan}, \citenamefont {Verberk},\ and\ \citenamefont
  {Schmidt}}]{Baranov2011}%
  \BibitemOpen
  \bibfield  {author} {\bibinfo {author} {\bibfnamefont {P.~G.}\ \bibnamefont
  {Baranov}}, \bibinfo {author} {\bibfnamefont {A.~P.}\ \bibnamefont
  {Bundakova}}, \bibinfo {author} {\bibfnamefont {A.~A.}\ \bibnamefont
  {Soltamova}}, \bibinfo {author} {\bibfnamefont {S.~B.}\ \bibnamefont
  {Orlinskii}}, \bibinfo {author} {\bibfnamefont {I.~V.}\ \bibnamefont
  {Borovykh}}, \bibinfo {author} {\bibfnamefont {R.}~\bibnamefont {Zondervan}},
  \bibinfo {author} {\bibfnamefont {R.}~\bibnamefont {Verberk}}, \ and\
  \bibinfo {author} {\bibfnamefont {J.}~\bibnamefont {Schmidt}},\ }\href
  {http://link.aps.org/doi/10.1103/PhysRevB.83.125203} {\bibfield  {journal}
  {\bibinfo  {journal} {Physical Review B}\ }\textbf {\bibinfo {volume} {83}},\
  \bibinfo {pages} {125203} (\bibinfo {year} {2011})}\BibitemShut {NoStop}%
\bibitem [{\citenamefont {Mizuochi}\ \emph {et~al.}(2002)\citenamefont
  {Mizuochi}, \citenamefont {Yamasaki}, \citenamefont {Takizawa}, \citenamefont
  {Morishita}, \citenamefont {Ohshima}, \citenamefont {Itoh},\ and\
  \citenamefont {Isoya}}]{Mizuochi2002}%
  \BibitemOpen
  \bibfield  {author} {\bibinfo {author} {\bibfnamefont {N.}~\bibnamefont
  {Mizuochi}}, \bibinfo {author} {\bibfnamefont {S.}~\bibnamefont {Yamasaki}},
  \bibinfo {author} {\bibfnamefont {H.}~\bibnamefont {Takizawa}}, \bibinfo
  {author} {\bibfnamefont {N.}~\bibnamefont {Morishita}}, \bibinfo {author}
  {\bibfnamefont {T.}~\bibnamefont {Ohshima}}, \bibinfo {author} {\bibfnamefont
  {H.}~\bibnamefont {Itoh}}, \ and\ \bibinfo {author} {\bibfnamefont
  {J.}~\bibnamefont {Isoya}},\ }\href
  {http://link.aps.org/doi/10.1103/PhysRevB.66.235202} {\bibfield  {journal}
  {\bibinfo  {journal} {Physical Review B}\ }\textbf {\bibinfo {volume} {66}},\
  \bibinfo {pages} {235202} (\bibinfo {year} {2002})}\BibitemShut {NoStop}%
\bibitem [{\citenamefont {Kraus}\ \emph
  {et~al.}(2014{\natexlab{a}})\citenamefont {Kraus}, \citenamefont {Soltamov},
  \citenamefont {Riedel}, \citenamefont {Vath}, \citenamefont {Fuchs},
  \citenamefont {Sperlich}, \citenamefont {Baranov}, \citenamefont {Dyakonov},\
  and\ \citenamefont {Astakhov}}]{Kraus2014}%
  \BibitemOpen
  \bibfield  {author} {\bibinfo {author} {\bibfnamefont {H.}~\bibnamefont
  {Kraus}}, \bibinfo {author} {\bibfnamefont {V.~A.}\ \bibnamefont {Soltamov}},
  \bibinfo {author} {\bibfnamefont {D.}~\bibnamefont {Riedel}}, \bibinfo
  {author} {\bibfnamefont {S.}~\bibnamefont {Vath}}, \bibinfo {author}
  {\bibfnamefont {F.}~\bibnamefont {Fuchs}}, \bibinfo {author} {\bibfnamefont
  {A.}~\bibnamefont {Sperlich}}, \bibinfo {author} {\bibfnamefont {P.~G.}\
  \bibnamefont {Baranov}}, \bibinfo {author} {\bibfnamefont {V.}~\bibnamefont
  {Dyakonov}}, \ and\ \bibinfo {author} {\bibfnamefont {G.~V.}\ \bibnamefont
  {Astakhov}},\ }\href {http://dx.doi.org/10.1038/nphys2826 10.1038/nphys2826
  http://www.nature.com/nphys/journal/vaop/ncurrent/abs/nphys2826.html\#supplementary-information}
  {\bibfield  {journal} {\bibinfo  {journal} {Nat Phys}\ }\textbf {\bibinfo
  {volume} {10}},\ \bibinfo {pages} {157} (\bibinfo {year}
  {2014}{\natexlab{a}})}\BibitemShut {NoStop}%
\bibitem [{\citenamefont {Sz\'{a}sz}\ \emph {et~al.}(2015)\citenamefont
  {Sz\'{a}sz}, \citenamefont {Iv\'{a}dy}, \citenamefont {Abrikosov},
  \citenamefont {Janz\'{e}n}, \citenamefont {Bockstedte},\ and\ \citenamefont
  {Gali}}]{Szasz2015}%
  \BibitemOpen
  \bibfield  {author} {\bibinfo {author} {\bibfnamefont {K.}~\bibnamefont
  {Sz\'{a}sz}}, \bibinfo {author} {\bibfnamefont {V.}~\bibnamefont
  {Iv\'{a}dy}}, \bibinfo {author} {\bibfnamefont {I.~A.}\ \bibnamefont
  {Abrikosov}}, \bibinfo {author} {\bibfnamefont {E.}~\bibnamefont
  {Janz\'{e}n}}, \bibinfo {author} {\bibfnamefont {M.}~\bibnamefont
  {Bockstedte}}, \ and\ \bibinfo {author} {\bibfnamefont {A.}~\bibnamefont
  {Gali}},\ }\href {http://link.aps.org/doi/10.1103/PhysRevB.91.121201}
  {\bibfield  {journal} {\bibinfo  {journal} {Physical Review B}\ }\textbf
  {\bibinfo {volume} {91}},\ \bibinfo {pages} {121201} (\bibinfo {year}
  {2015})}\BibitemShut {NoStop}%
\bibitem [{\citenamefont {Gali}\ \emph {et~al.}(2010)\citenamefont {Gali},
  \citenamefont {G\"{a}llstr\"{o}m}, \citenamefont {Son},\ and\ \citenamefont
  {Janz\'{e}n}}]{Gali2010}%
  \BibitemOpen
  \bibfield  {author} {\bibinfo {author} {\bibfnamefont {A.}~\bibnamefont
  {Gali}}, \bibinfo {author} {\bibfnamefont {A.}~\bibnamefont
  {G\"{a}llstr\"{o}m}}, \bibinfo {author} {\bibfnamefont {N.~T.}\ \bibnamefont
  {Son}}, \ and\ \bibinfo {author} {\bibfnamefont {E.}~\bibnamefont
  {Janz\'{e}n}},\ }in\ \href@noop {} {\emph {\bibinfo {booktitle} {Materials
  Science Forum}}},\ Vol.\ \bibinfo {volume} {645}\ (\bibinfo  {publisher}
  {Trans Tech Publ},\ \bibinfo {year} {2010})\ pp.\ \bibinfo {pages}
  {395--397}\BibitemShut {NoStop}%
\bibitem [{\citenamefont {Son}\ \emph {et~al.}(2006)\citenamefont {Son},
  \citenamefont {Carlsson}, \citenamefont {ul~Hassan}, \citenamefont
  {Janz\'{e}n}, \citenamefont {Umeda}, \citenamefont {Isoya}, \citenamefont
  {Gali}, \citenamefont {Bockstedte}, \citenamefont {Morishita}, \citenamefont
  {Ohshima},\ and\ \citenamefont {Itoh}}]{Son2006a}%
  \BibitemOpen
  \bibfield  {author} {\bibinfo {author} {\bibfnamefont {N.~T.}\ \bibnamefont
  {Son}}, \bibinfo {author} {\bibfnamefont {P.}~\bibnamefont {Carlsson}},
  \bibinfo {author} {\bibfnamefont {J.}~\bibnamefont {ul~Hassan}}, \bibinfo
  {author} {\bibfnamefont {E.}~\bibnamefont {Janz\'{e}n}}, \bibinfo {author}
  {\bibfnamefont {T.}~\bibnamefont {Umeda}}, \bibinfo {author} {\bibfnamefont
  {J.}~\bibnamefont {Isoya}}, \bibinfo {author} {\bibfnamefont
  {A.}~\bibnamefont {Gali}}, \bibinfo {author} {\bibfnamefont {M.}~\bibnamefont
  {Bockstedte}}, \bibinfo {author} {\bibfnamefont {N.}~\bibnamefont
  {Morishita}}, \bibinfo {author} {\bibfnamefont {T.}~\bibnamefont {Ohshima}},
  \ and\ \bibinfo {author} {\bibfnamefont {H.}~\bibnamefont {Itoh}},\ }\href
  {http://link.aps.org/doi/10.1103/PhysRevLett.96.055501} {\bibfield  {journal}
  {\bibinfo  {journal} {Physical Review Letters}\ }\textbf {\bibinfo {volume}
  {96}},\ \bibinfo {pages} {55501} (\bibinfo {year} {2006})}\BibitemShut
  {NoStop}%
\bibitem [{\citenamefont {Stevenson}(1984)}]{Stevenson1984}%
  \BibitemOpen
  \bibfield  {author} {\bibinfo {author} {\bibfnamefont {R.~C.}\ \bibnamefont
  {Stevenson}},\ }\href {\doibase 10.1016/0022-2364(84)90231-2} {\bibfield
  {journal} {\bibinfo  {journal} {Journal of Magnetic Resonance (1969)}\
  }\textbf {\bibinfo {volume} {57}},\ \bibinfo {pages} {24} (\bibinfo {year}
  {1984})}\BibitemShut {NoStop}%
\bibitem [{\citenamefont {Atherton}(1993)}]{Atherton1993}%
  \BibitemOpen
  \bibfield  {author} {\bibinfo {author} {\bibfnamefont {N.~M.}\ \bibnamefont
  {Atherton}},\ }\href@noop {} {\emph {\bibinfo {title} {Ellis Horwood series
  in physical chemistry}}}\ (\bibinfo  {publisher} {Ellis Horwood},\ \bibinfo
  {address} {Chichester},\ \bibinfo {year} {1993})\BibitemShut {NoStop}%
\bibitem [{\citenamefont {Balasubramanian}\ \emph {et~al.}(2008)\citenamefont
  {Balasubramanian}, \citenamefont {Chan}, \citenamefont {Kolesov},
  \citenamefont {Al-Hmoud}, \citenamefont {Tisler}, \citenamefont {Shin},
  \citenamefont {Kim}, \citenamefont {Wojcik}, \citenamefont {Hemmer},
  \citenamefont {Krueger}, \citenamefont {Hanke}, \citenamefont
  {Leitenstorfer}, \citenamefont {Bratschitsch}, \citenamefont {Jelezko},\ and\
  \citenamefont {Wrachtrup}}]{Balasubramanian2008}%
  \BibitemOpen
  \bibfield  {author} {\bibinfo {author} {\bibfnamefont {G.}~\bibnamefont
  {Balasubramanian}}, \bibinfo {author} {\bibfnamefont {I.~Y.}\ \bibnamefont
  {Chan}}, \bibinfo {author} {\bibfnamefont {R.}~\bibnamefont {Kolesov}},
  \bibinfo {author} {\bibfnamefont {M.}~\bibnamefont {Al-Hmoud}}, \bibinfo
  {author} {\bibfnamefont {J.}~\bibnamefont {Tisler}}, \bibinfo {author}
  {\bibfnamefont {C.}~\bibnamefont {Shin}}, \bibinfo {author} {\bibfnamefont
  {C.}~\bibnamefont {Kim}}, \bibinfo {author} {\bibfnamefont {A.}~\bibnamefont
  {Wojcik}}, \bibinfo {author} {\bibfnamefont {P.~R.}\ \bibnamefont {Hemmer}},
  \bibinfo {author} {\bibfnamefont {A.}~\bibnamefont {Krueger}}, \bibinfo
  {author} {\bibfnamefont {T.}~\bibnamefont {Hanke}}, \bibinfo {author}
  {\bibfnamefont {A.}~\bibnamefont {Leitenstorfer}}, \bibinfo {author}
  {\bibfnamefont {R.}~\bibnamefont {Bratschitsch}}, \bibinfo {author}
  {\bibfnamefont {F.}~\bibnamefont {Jelezko}}, \ and\ \bibinfo {author}
  {\bibfnamefont {J.}~\bibnamefont {Wrachtrup}},\ }\href {\doibase
  http://www.nature.com/nature/journal/v455/n7213/suppinfo/nature07278\_S1.html}
  {\bibfield  {journal} {\bibinfo  {journal} {Nature}\ }\textbf {\bibinfo
  {volume} {455}},\ \bibinfo {pages} {648} (\bibinfo {year}
  {2008})}\BibitemShut {NoStop}%
\bibitem [{\citenamefont {Steinert}(2010)}]{Steinert2010}%
  \BibitemOpen
  \bibfield  {author} {\bibinfo {author} {\bibfnamefont {S.}~\bibnamefont
  {Steinert}},\ }\href {http://dx.doi.org/doi/10.1063/1.3385689} {\bibfield
  {journal} {\bibinfo  {journal} {Rev. Sci. Instrum.}\ }\textbf {\bibinfo
  {volume} {81}},\ \bibinfo {pages} {43705} (\bibinfo {year}
  {2010})}\BibitemShut {NoStop}%
\bibitem [{\citenamefont {Degen}(2008)}]{Degen2008}%
  \BibitemOpen
  \bibfield  {author} {\bibinfo {author} {\bibfnamefont {C.~L.}\ \bibnamefont
  {Degen}},\ }\href {http://dx.doi.org/10.1063/1.2943282} {\bibfield  {journal}
  {\bibinfo  {journal} {Applied Physics Letters}\ }\textbf {\bibinfo {volume}
  {92}},\ \bibinfo {pages} {243111} (\bibinfo {year} {2008})}\BibitemShut
  {NoStop}%
\bibitem [{\citenamefont {Taylor}\ \emph {et~al.}(2008)\citenamefont {Taylor},
  \citenamefont {Cappellaro}, \citenamefont {Childress}, \citenamefont {Jiang},
  \citenamefont {Budker}, \citenamefont {Hemmer}, \citenamefont {Yacoby},
  \citenamefont {Walsworth},\ and\ \citenamefont {Lukin}}]{Taylor2008}%
  \BibitemOpen
  \bibfield  {author} {\bibinfo {author} {\bibfnamefont {J.~M.}\ \bibnamefont
  {Taylor}}, \bibinfo {author} {\bibfnamefont {P.}~\bibnamefont {Cappellaro}},
  \bibinfo {author} {\bibfnamefont {L.}~\bibnamefont {Childress}}, \bibinfo
  {author} {\bibfnamefont {L.}~\bibnamefont {Jiang}}, \bibinfo {author}
  {\bibfnamefont {D.}~\bibnamefont {Budker}}, \bibinfo {author} {\bibfnamefont
  {P.~R.}\ \bibnamefont {Hemmer}}, \bibinfo {author} {\bibfnamefont
  {A.}~\bibnamefont {Yacoby}}, \bibinfo {author} {\bibfnamefont
  {R.}~\bibnamefont {Walsworth}}, \ and\ \bibinfo {author} {\bibfnamefont
  {M.~D.}\ \bibnamefont {Lukin}},\ }\href {http://dx.doi.org/10.1038/nphys1075}
  {\bibfield  {journal} {\bibinfo  {journal} {Nat Phys}\ }\textbf {\bibinfo
  {volume} {4}},\ \bibinfo {pages} {810} (\bibinfo {year} {2008})}\BibitemShut
  {NoStop}%
\bibitem [{\citenamefont {Clevenson}\ \emph {et~al.}(2015)\citenamefont
  {Clevenson}, \citenamefont {Trusheim}, \citenamefont {Teale}, \citenamefont
  {Schroder}, \citenamefont {Braje},\ and\ \citenamefont
  {Englund}}]{Clevenson2015}%
  \BibitemOpen
  \bibfield  {author} {\bibinfo {author} {\bibfnamefont {H.}~\bibnamefont
  {Clevenson}}, \bibinfo {author} {\bibfnamefont {M.~E.}\ \bibnamefont
  {Trusheim}}, \bibinfo {author} {\bibfnamefont {C.}~\bibnamefont {Teale}},
  \bibinfo {author} {\bibfnamefont {T.}~\bibnamefont {Schroder}}, \bibinfo
  {author} {\bibfnamefont {D.}~\bibnamefont {Braje}}, \ and\ \bibinfo {author}
  {\bibfnamefont {D.}~\bibnamefont {Englund}},\ }\href
  {http://dx.doi.org/10.1038/nphys3291 10.1038/nphys3291
  http://www.nature.com/nphys/journal/v11/n5/abs/nphys3291.html\#supplementary-information}
  {\bibfield  {journal} {\bibinfo  {journal} {Nat Phys}\ }\textbf {\bibinfo
  {volume} {11}},\ \bibinfo {pages} {393} (\bibinfo {year} {2015})}\BibitemShut
  {NoStop}%
\bibitem [{\citenamefont {Balasubramanian}\ \emph {et~al.}(2009)\citenamefont
  {Balasubramanian}, \citenamefont {Neumann}, \citenamefont {Twitchen},
  \citenamefont {Markham}, \citenamefont {Kolesov}, \citenamefont {Mizuochi},
  \citenamefont {Isoya}, \citenamefont {Achard}, \citenamefont {Beck},
  \citenamefont {Tissler}, \citenamefont {Jacques}, \citenamefont {Hemmer},
  \citenamefont {Jelezko},\ and\ \citenamefont
  {Wrachtrup}}]{Balasubramanian2009a}%
  \BibitemOpen
  \bibfield  {author} {\bibinfo {author} {\bibfnamefont {G.}~\bibnamefont
  {Balasubramanian}}, \bibinfo {author} {\bibfnamefont {P.}~\bibnamefont
  {Neumann}}, \bibinfo {author} {\bibfnamefont {D.}~\bibnamefont {Twitchen}},
  \bibinfo {author} {\bibfnamefont {M.}~\bibnamefont {Markham}}, \bibinfo
  {author} {\bibfnamefont {R.}~\bibnamefont {Kolesov}}, \bibinfo {author}
  {\bibfnamefont {N.}~\bibnamefont {Mizuochi}}, \bibinfo {author}
  {\bibfnamefont {J.}~\bibnamefont {Isoya}}, \bibinfo {author} {\bibfnamefont
  {J.}~\bibnamefont {Achard}}, \bibinfo {author} {\bibfnamefont
  {J.}~\bibnamefont {Beck}}, \bibinfo {author} {\bibfnamefont {J.}~\bibnamefont
  {Tissler}}, \bibinfo {author} {\bibfnamefont {V.}~\bibnamefont {Jacques}},
  \bibinfo {author} {\bibfnamefont {P.~R.}\ \bibnamefont {Hemmer}}, \bibinfo
  {author} {\bibfnamefont {F.}~\bibnamefont {Jelezko}}, \ and\ \bibinfo
  {author} {\bibfnamefont {J.}~\bibnamefont {Wrachtrup}},\ }\href
  {http://dx.doi.org/10.1038/nmat2420} {\bibfield  {journal} {\bibinfo
  {journal} {Nat Mater}\ }\textbf {\bibinfo {volume} {8}},\ \bibinfo {pages}
  {383} (\bibinfo {year} {2009})}\BibitemShut {NoStop}%
\bibitem [{\citenamefont {Wolf}\ \emph {et~al.}(2014)\citenamefont {Wolf},
  \citenamefont {Neumann}, \citenamefont {Isoya},\ and\ \citenamefont
  {Wrachtrup}}]{2014arXiv1411.6553W}%
  \BibitemOpen
  \bibfield  {author} {\bibinfo {author} {\bibfnamefont {T.}~\bibnamefont
  {Wolf}}, \bibinfo {author} {\bibfnamefont {P.}~\bibnamefont {Neumann}},
  \bibinfo {author} {\bibfnamefont {J.}~\bibnamefont {Isoya}}, \ and\ \bibinfo
  {author} {\bibfnamefont {J.}~\bibnamefont {Wrachtrup}},\ }\href@noop {}
  {\bibfield  {journal} {\bibinfo  {journal} {ArXiv e-prints}\ } (\bibinfo
  {year} {2014})},\ \Eprint {http://arxiv.org/abs/1411.6553} {arXiv:1411.6553
  [quant-ph]} \BibitemShut {NoStop}%
\bibitem [{\citenamefont {Mizuochi}\ \emph {et~al.}(2003)\citenamefont
  {Mizuochi}, \citenamefont {Yamasaki}, \citenamefont {Takizawa}, \citenamefont
  {Morishita}, \citenamefont {Ohshima}, \citenamefont {Itoh},\ and\
  \citenamefont {Isoya}}]{Mizuochi2003}%
  \BibitemOpen
  \bibfield  {author} {\bibinfo {author} {\bibfnamefont {N.}~\bibnamefont
  {Mizuochi}}, \bibinfo {author} {\bibfnamefont {S.}~\bibnamefont {Yamasaki}},
  \bibinfo {author} {\bibfnamefont {H.}~\bibnamefont {Takizawa}}, \bibinfo
  {author} {\bibfnamefont {N.}~\bibnamefont {Morishita}}, \bibinfo {author}
  {\bibfnamefont {T.}~\bibnamefont {Ohshima}}, \bibinfo {author} {\bibfnamefont
  {H.}~\bibnamefont {Itoh}}, \ and\ \bibinfo {author} {\bibfnamefont
  {J.}~\bibnamefont {Isoya}},\ }\href
  {http://link.aps.org/doi/10.1103/PhysRevB.68.165206} {\bibfield  {journal}
  {\bibinfo  {journal} {Physical Review B}\ }\textbf {\bibinfo {volume} {68}},\
  \bibinfo {pages} {165206} (\bibinfo {year} {2003})}\BibitemShut {NoStop}%
\bibitem [{\citenamefont {Isoya}\ \emph {et~al.}(2008)\citenamefont {Isoya},
  \citenamefont {Umeda}, \citenamefont {Mizuochi}, \citenamefont {Son},
  \citenamefont {Janz\'{e}n},\ and\ \citenamefont {Ohshima}}]{Isoya2008}%
  \BibitemOpen
  \bibfield  {author} {\bibinfo {author} {\bibfnamefont {J.}~\bibnamefont
  {Isoya}}, \bibinfo {author} {\bibfnamefont {T.}~\bibnamefont {Umeda}},
  \bibinfo {author} {\bibfnamefont {N.}~\bibnamefont {Mizuochi}}, \bibinfo
  {author} {\bibfnamefont {N.~T.}\ \bibnamefont {Son}}, \bibinfo {author}
  {\bibfnamefont {E.}~\bibnamefont {Janz\'{e}n}}, \ and\ \bibinfo {author}
  {\bibfnamefont {T.}~\bibnamefont {Ohshima}},\ }\href {\doibase
  10.1002/pssb.200844209} {\bibfield  {journal} {\bibinfo  {journal} {physica
  status solidi (b)}\ }\textbf {\bibinfo {volume} {245}},\ \bibinfo {pages}
  {1298} (\bibinfo {year} {2008})}\BibitemShut {NoStop}%
\bibitem [{\citenamefont {Mizuochi}\ \emph {et~al.}(2005)\citenamefont
  {Mizuochi}, \citenamefont {Yamasaki}, \citenamefont {Takizawa}, \citenamefont
  {Morishita}, \citenamefont {Ohshima}, \citenamefont {Itoh}, \citenamefont
  {Umeda},\ and\ \citenamefont {Isoya}}]{Mizuochi2005prb}%
  \BibitemOpen
  \bibfield  {author} {\bibinfo {author} {\bibfnamefont {N.}~\bibnamefont
  {Mizuochi}}, \bibinfo {author} {\bibfnamefont {S.}~\bibnamefont {Yamasaki}},
  \bibinfo {author} {\bibfnamefont {H.}~\bibnamefont {Takizawa}}, \bibinfo
  {author} {\bibfnamefont {N.}~\bibnamefont {Morishita}}, \bibinfo {author}
  {\bibfnamefont {T.}~\bibnamefont {Ohshima}}, \bibinfo {author} {\bibfnamefont
  {H.}~\bibnamefont {Itoh}}, \bibinfo {author} {\bibfnamefont {T.}~\bibnamefont
  {Umeda}}, \ and\ \bibinfo {author} {\bibfnamefont {J.}~\bibnamefont
  {Isoya}},\ }\href {http://link.aps.org/doi/10.1103/PhysRevB.72.235208}
  {\bibfield  {journal} {\bibinfo  {journal} {Physical Review B}\ }\textbf
  {\bibinfo {volume} {72}},\ \bibinfo {pages} {235208} (\bibinfo {year}
  {2005})}\BibitemShut {NoStop}%
\bibitem [{\citenamefont {Soltamov}\ \emph {et~al.}(2012)\citenamefont
  {Soltamov}, \citenamefont {Soltamova}, \citenamefont {Baranov},\ and\
  \citenamefont {Proskuryakov}}]{Soltamov2012}%
  \BibitemOpen
  \bibfield  {author} {\bibinfo {author} {\bibfnamefont {V.~A.}\ \bibnamefont
  {Soltamov}}, \bibinfo {author} {\bibfnamefont {A.~A.}\ \bibnamefont
  {Soltamova}}, \bibinfo {author} {\bibfnamefont {P.~G.}\ \bibnamefont
  {Baranov}}, \ and\ \bibinfo {author} {\bibfnamefont {I.~I.}\ \bibnamefont
  {Proskuryakov}},\ }\href
  {http://link.aps.org/doi/10.1103/PhysRevLett.108.226402} {\bibfield
  {journal} {\bibinfo  {journal} {Physical Review Letters}\ }\textbf {\bibinfo
  {volume} {108}},\ \bibinfo {pages} {226402} (\bibinfo {year}
  {2012})}\BibitemShut {NoStop}%
\bibitem [{\citenamefont {Simin}\ \emph {et~al.}(2015)\citenamefont {Simin},
  \citenamefont {Fuchs}, \citenamefont {Kraus}, \citenamefont {Sperlich},
  \citenamefont {Baranov}, \citenamefont {Astakhov},\ and\ \citenamefont
  {Dyakonov}}]{Simin2015}%
  \BibitemOpen
  \bibfield  {author} {\bibinfo {author} {\bibfnamefont {D.}~\bibnamefont
  {Simin}}, \bibinfo {author} {\bibfnamefont {F.}~\bibnamefont {Fuchs}},
  \bibinfo {author} {\bibfnamefont {H.}~\bibnamefont {Kraus}}, \bibinfo
  {author} {\bibfnamefont {A.}~\bibnamefont {Sperlich}}, \bibinfo {author}
  {\bibfnamefont {P.~G.}\ \bibnamefont {Baranov}}, \bibinfo {author}
  {\bibfnamefont {G.~V.}\ \bibnamefont {Astakhov}}, \ and\ \bibinfo {author}
  {\bibfnamefont {V.}~\bibnamefont {Dyakonov}},\ }\href
  {http://link.aps.org/doi/10.1103/PhysRevApplied.4.014009} {\bibfield
  {journal} {\bibinfo  {journal} {Physical Review Applied}\ }\textbf {\bibinfo
  {volume} {4}},\ \bibinfo {pages} {14009} (\bibinfo {year} {2015})},\ \Eprint
  {http://arxiv.org/abs/1505.00176} {arXiv:1505.00176 [cond-mat.mtrl-sci]}
  \BibitemShut {NoStop}%
\bibitem [{\citenamefont {Kraus}\ \emph
  {et~al.}(2014{\natexlab{b}})\citenamefont {Kraus}, \citenamefont {Soltamov},
  \citenamefont {Fuchs}, \citenamefont {Simin}, \citenamefont {Sperlich},
  \citenamefont {Baranov}, \citenamefont {Astakhov},\ and\ \citenamefont
  {Dyakonov}}]{Kraus2014a}%
  \BibitemOpen
  \bibfield  {author} {\bibinfo {author} {\bibfnamefont {H.}~\bibnamefont
  {Kraus}}, \bibinfo {author} {\bibfnamefont {V.~A.}\ \bibnamefont {Soltamov}},
  \bibinfo {author} {\bibfnamefont {F.}~\bibnamefont {Fuchs}}, \bibinfo
  {author} {\bibfnamefont {D.}~\bibnamefont {Simin}}, \bibinfo {author}
  {\bibfnamefont {A.}~\bibnamefont {Sperlich}}, \bibinfo {author}
  {\bibfnamefont {P.~G.}\ \bibnamefont {Baranov}}, \bibinfo {author}
  {\bibfnamefont {G.~V.}\ \bibnamefont {Astakhov}}, \ and\ \bibinfo {author}
  {\bibfnamefont {V.}~\bibnamefont {Dyakonov}},\ }\href
  {http://dx.doi.org/10.1038/srep05303 10.1038/srep05303
  http://www.nature.com/srep/2014/140704/srep05303/abs/srep05303.html\#supplementary-information}
  {\bibfield  {journal} {\bibinfo  {journal} {Sci. Rep.}\ }\textbf {\bibinfo
  {volume} {4}} (\bibinfo {year} {2014}{\natexlab{b}})}\BibitemShut {NoStop}%
\bibitem [{\citenamefont {Morton}\ \emph {et~al.}(2005)\citenamefont {Morton},
  \citenamefont {Tyryshkin}, \citenamefont {Ardavan}, \citenamefont
  {Porfyrakis}, \citenamefont {Lyon},\ and\ \citenamefont
  {Briggs}}]{Morton2005JCP}%
  \BibitemOpen
  \bibfield  {author} {\bibinfo {author} {\bibfnamefont {J.~J.~L.}\
  \bibnamefont {Morton}}, \bibinfo {author} {\bibfnamefont {A.~M.}\
  \bibnamefont {Tyryshkin}}, \bibinfo {author} {\bibfnamefont {A.}~\bibnamefont
  {Ardavan}}, \bibinfo {author} {\bibfnamefont {K.}~\bibnamefont {Porfyrakis}},
  \bibinfo {author} {\bibfnamefont {S.~A.}\ \bibnamefont {Lyon}}, \ and\
  \bibinfo {author} {\bibfnamefont {G.~A.~D.}\ \bibnamefont {Briggs}},\ }\href
  {\doibase http://dx.doi.org/10.1063/1.1888585} {\bibfield  {journal}
  {\bibinfo  {journal} {The Journal of Chemical Physics}\ }\textbf {\bibinfo
  {volume} {122}},\  (\bibinfo {year} {2005})}\BibitemShut {NoStop}%
\bibitem [{\citenamefont {Knapp}\ \emph {et~al.}(1998)\citenamefont {Knapp},
  \citenamefont {Weiden}, \citenamefont {Kass}, \citenamefont {Dinse},
  \citenamefont {Pietzak}, \citenamefont {Waiblinger},\ and\ \citenamefont
  {Weidinger}}]{Knapp1998}%
  \BibitemOpen
  \bibfield  {author} {\bibinfo {author} {\bibfnamefont {C.}~\bibnamefont
  {Knapp}}, \bibinfo {author} {\bibfnamefont {N.}~\bibnamefont {Weiden}},
  \bibinfo {author} {\bibfnamefont {H.}~\bibnamefont {Kass}}, \bibinfo {author}
  {\bibfnamefont {K.-P.}\ \bibnamefont {Dinse}}, \bibinfo {author}
  {\bibfnamefont {B.}~\bibnamefont {Pietzak}}, \bibinfo {author} {\bibfnamefont
  {M.}~\bibnamefont {Waiblinger}}, \ and\ \bibinfo {author} {\bibfnamefont
  {A.}~\bibnamefont {Weidinger}},\ }\href {\doibase 10.1080/00268979809483233}
  {\bibfield  {journal} {\bibinfo  {journal} {Molecular Physics}\ }\textbf
  {\bibinfo {volume} {95}},\ \bibinfo {pages} {999} (\bibinfo {year}
  {1998})}\BibitemShut {NoStop}%
\bibitem [{\citenamefont {Harneit}(2002)}]{Harneit2002pra}%
  \BibitemOpen
  \bibfield  {author} {\bibinfo {author} {\bibfnamefont {W.}~\bibnamefont
  {Harneit}},\ }\href {http://link.aps.org/doi/10.1103/PhysRevA.65.032322}
  {\bibfield  {journal} {\bibinfo  {journal} {Physical Review A}\ }\textbf
  {\bibinfo {volume} {65}},\ \bibinfo {pages} {32322} (\bibinfo {year}
  {2002})}\BibitemShut {NoStop}%
\bibitem [{\citenamefont {Benjamin}\ \emph {et~al.}(2006)\citenamefont
  {Benjamin}, \citenamefont {Ardavan}, \citenamefont {Briggs}, \citenamefont
  {Britz}, \citenamefont {Gunlycke}, \citenamefont {Jefferson}, \citenamefont
  {Jones}, \citenamefont {Leigh}, \citenamefont {Lovett},\ and\ \citenamefont
  {Khlobystov}}]{Benjamin2006JP}%
  \BibitemOpen
  \bibfield  {author} {\bibinfo {author} {\bibfnamefont {S.~C.}\ \bibnamefont
  {Benjamin}}, \bibinfo {author} {\bibfnamefont {A.}~\bibnamefont {Ardavan}},
  \bibinfo {author} {\bibfnamefont {G.~A.~D.}\ \bibnamefont {Briggs}}, \bibinfo
  {author} {\bibfnamefont {D.~A.}\ \bibnamefont {Britz}}, \bibinfo {author}
  {\bibfnamefont {D.}~\bibnamefont {Gunlycke}}, \bibinfo {author}
  {\bibfnamefont {J.}~\bibnamefont {Jefferson}}, \bibinfo {author}
  {\bibfnamefont {M.~A.~G.}\ \bibnamefont {Jones}}, \bibinfo {author}
  {\bibfnamefont {D.~F.}\ \bibnamefont {Leigh}}, \bibinfo {author}
  {\bibfnamefont {B.~W.}\ \bibnamefont {Lovett}}, \ and\ \bibinfo {author}
  {\bibfnamefont {A.~N.}\ \bibnamefont {Khlobystov}},\ }\href@noop {}
  {\bibfield  {journal} {\bibinfo  {journal} {Journal of Physics: Condensed
  Matter}\ }\textbf {\bibinfo {volume} {18}},\ \bibinfo {pages} {S867}
  (\bibinfo {year} {2006})}\BibitemShut {NoStop}%
\bibitem [{\citenamefont {Mizuochi}\ \emph {et~al.}(1999)\citenamefont
  {Mizuochi}, \citenamefont {Ohba},\ and\ \citenamefont
  {Yamauchi}}]{MizuochiJCP1999}%
  \BibitemOpen
  \bibfield  {author} {\bibinfo {author} {\bibfnamefont {N.}~\bibnamefont
  {Mizuochi}}, \bibinfo {author} {\bibfnamefont {Y.}~\bibnamefont {Ohba}}, \
  and\ \bibinfo {author} {\bibfnamefont {S.}~\bibnamefont {Yamauchi}},\
  }\href@noop {} {\bibfield  {journal} {\bibinfo  {journal} {The Journal of
  Chemical Physics}\ }\textbf {\bibinfo {volume} {111}} (\bibinfo {year}
  {1999})}\BibitemShut {NoStop}%
\bibitem [{\citenamefont {Teki}\ \emph {et~al.}(2001)\citenamefont {Teki},
  \citenamefont {Miyamoto}, \citenamefont {Nakatsuji},\ and\ \citenamefont
  {Miura}}]{Teki2001JACS}%
  \BibitemOpen
  \bibfield  {author} {\bibinfo {author} {\bibfnamefont {Y.}~\bibnamefont
  {Teki}}, \bibinfo {author} {\bibfnamefont {S.}~\bibnamefont {Miyamoto}},
  \bibinfo {author} {\bibfnamefont {M.}~\bibnamefont {Nakatsuji}}, \ and\
  \bibinfo {author} {\bibfnamefont {Y.}~\bibnamefont {Miura}},\ }\href
  {\doibase 10.1021/ja001920k} {\bibfield  {journal} {\bibinfo  {journal}
  {Journal of the American Chemical Society}\ }\textbf {\bibinfo {volume}
  {123}},\ \bibinfo {pages} {294} (\bibinfo {year} {2001})}\BibitemShut
  {NoStop}%
\bibitem [{\citenamefont {Kothe}\ \emph {et~al.}(1980)\citenamefont {Kothe},
  \citenamefont {Kim},\ and\ \citenamefont {Weissman}}]{Kothe1980cpl}%
  \BibitemOpen
  \bibfield  {author} {\bibinfo {author} {\bibfnamefont {G.}~\bibnamefont
  {Kothe}}, \bibinfo {author} {\bibfnamefont {S.~S.}\ \bibnamefont {Kim}}, \
  and\ \bibinfo {author} {\bibfnamefont {S.~I.}\ \bibnamefont {Weissman}},\
  }\href {\doibase http://dx.doi.org/10.1016/0009-2614(80)80199-0} {\bibfield
  {journal} {\bibinfo  {journal} {Chemical Physics Letters}\ }\textbf {\bibinfo
  {volume} {71}},\ \bibinfo {pages} {445} (\bibinfo {year} {1980})}\BibitemShut
  {NoStop}%
\bibitem [{\citenamefont {Teki}\ \emph {et~al.}(2008)\citenamefont {Teki},
  \citenamefont {Tamekuni}, \citenamefont {Haruta}, \citenamefont {Takeuchi},\
  and\ \citenamefont {Miura}}]{Teki2008}%
  \BibitemOpen
  \bibfield  {author} {\bibinfo {author} {\bibfnamefont {Y.}~\bibnamefont
  {Teki}}, \bibinfo {author} {\bibfnamefont {H.}~\bibnamefont {Tamekuni}},
  \bibinfo {author} {\bibfnamefont {K.}~\bibnamefont {Haruta}}, \bibinfo
  {author} {\bibfnamefont {J.}~\bibnamefont {Takeuchi}}, \ and\ \bibinfo
  {author} {\bibfnamefont {Y.}~\bibnamefont {Miura}},\ }\href {\doibase
  10.1039/B714868B} {\bibfield  {journal} {\bibinfo  {journal} {Journal of
  Materials Chemistry}\ }\textbf {\bibinfo {volume} {18}},\ \bibinfo {pages}
  {381} (\bibinfo {year} {2008})}\BibitemShut {NoStop}%
\bibitem [{\citenamefont {Isoya}\ \emph {et~al.}(1990)\citenamefont {Isoya},
  \citenamefont {Kanda}, \citenamefont {Norris}, \citenamefont {Tang},\ and\
  \citenamefont {Bowman}}]{Isoya1990}%
  \BibitemOpen
  \bibfield  {author} {\bibinfo {author} {\bibfnamefont {J.}~\bibnamefont
  {Isoya}}, \bibinfo {author} {\bibfnamefont {H.}~\bibnamefont {Kanda}},
  \bibinfo {author} {\bibfnamefont {J.~R.}\ \bibnamefont {Norris}}, \bibinfo
  {author} {\bibfnamefont {J.}~\bibnamefont {Tang}}, \ and\ \bibinfo {author}
  {\bibfnamefont {M.~K.}\ \bibnamefont {Bowman}},\ }\href
  {http://link.aps.org/doi/10.1103/PhysRevB.41.3905} {\bibfield  {journal}
  {\bibinfo  {journal} {Physical Review B}\ }\textbf {\bibinfo {volume} {41}},\
  \bibinfo {pages} {3905} (\bibinfo {year} {1990})}\BibitemShut {NoStop}%
\bibitem [{\citenamefont {van Leeuwen}\ \emph {et~al.}(1986)\citenamefont {van
  Leeuwen}, \citenamefont {Vreeker},\ and\ \citenamefont
  {Glasbeek}}]{vanLeeuwen19860rb}%
  \BibitemOpen
  \bibfield  {author} {\bibinfo {author} {\bibfnamefont {P.~A.}\ \bibnamefont
  {van Leeuwen}}, \bibinfo {author} {\bibfnamefont {R.}~\bibnamefont
  {Vreeker}}, \ and\ \bibinfo {author} {\bibfnamefont {M.}~\bibnamefont
  {Glasbeek}},\ }\href {http://link.aps.org/doi/10.1103/PhysRevB.34.3483}
  {\bibfield  {journal} {\bibinfo  {journal} {Physical Review B}\ }\textbf
  {\bibinfo {volume} {34}},\ \bibinfo {pages} {3483} (\bibinfo {year}
  {1986})}\BibitemShut {NoStop}%
\bibitem [{\citenamefont {de~Groot}\ and\ \citenamefont {van~der
  Waals}(1960)}]{DeGroot1960}%
  \BibitemOpen
  \bibfield  {author} {\bibinfo {author} {\bibfnamefont {M.~S.}\ \bibnamefont
  {de~Groot}}\ and\ \bibinfo {author} {\bibfnamefont {J.~H.}\ \bibnamefont
  {van~der Waals}},\ }\href {\doibase 10.1080/00268976000100221} {\bibfield
  {journal} {\bibinfo  {journal} {Molecular Physics}\ }\textbf {\bibinfo
  {volume} {3}},\ \bibinfo {pages} {190} (\bibinfo {year} {1960})}\BibitemShut
  {NoStop}%
\bibitem [{\citenamefont {He}\ \emph {et~al.}(1993)\citenamefont {He},
  \citenamefont {Manson},\ and\ \citenamefont {Fisk}}]{He1993}%
  \BibitemOpen
  \bibfield  {author} {\bibinfo {author} {\bibfnamefont {X.-F.}\ \bibnamefont
  {He}}, \bibinfo {author} {\bibfnamefont {N.~B.}\ \bibnamefont {Manson}}, \
  and\ \bibinfo {author} {\bibfnamefont {P.~T.~H.}\ \bibnamefont {Fisk}},\
  }\href {http://link.aps.org/doi/10.1103/PhysRevB.47.8809} {\bibfield
  {journal} {\bibinfo  {journal} {Physical Review B}\ }\textbf {\bibinfo
  {volume} {47}},\ \bibinfo {pages} {8809} (\bibinfo {year}
  {1993})}\BibitemShut {NoStop}%
\bibitem [{\citenamefont {Dolde}\ \emph {et~al.}(2011)\citenamefont {Dolde},
  \citenamefont {Fedder}, \citenamefont {Doherty}, \citenamefont {Nobauer},
  \citenamefont {Rempp}, \citenamefont {Balasubramanian}, \citenamefont {Wolf},
  \citenamefont {Reinhard}, \citenamefont {Hollenberg}, \citenamefont
  {Jelezko},\ and\ \citenamefont {Wrachtrup}}]{Dolde2011}%
  \BibitemOpen
  \bibfield  {author} {\bibinfo {author} {\bibfnamefont {F.}~\bibnamefont
  {Dolde}}, \bibinfo {author} {\bibfnamefont {H.}~\bibnamefont {Fedder}},
  \bibinfo {author} {\bibfnamefont {M.~W.}\ \bibnamefont {Doherty}}, \bibinfo
  {author} {\bibfnamefont {T.}~\bibnamefont {Nobauer}}, \bibinfo {author}
  {\bibfnamefont {F.}~\bibnamefont {Rempp}}, \bibinfo {author} {\bibfnamefont
  {G.}~\bibnamefont {Balasubramanian}}, \bibinfo {author} {\bibfnamefont
  {T.}~\bibnamefont {Wolf}}, \bibinfo {author} {\bibfnamefont {F.}~\bibnamefont
  {Reinhard}}, \bibinfo {author} {\bibfnamefont {L.~C.~L.}\ \bibnamefont
  {Hollenberg}}, \bibinfo {author} {\bibfnamefont {F.}~\bibnamefont {Jelezko}},
  \ and\ \bibinfo {author} {\bibfnamefont {J.}~\bibnamefont {Wrachtrup}},\
  }\href {http://dx.doi.org/10.1038/nphys1969
  http://www.nature.com/nphys/journal/v7/n6/abs/nphys1969.html\#supplementary-information}
  {\bibfield  {journal} {\bibinfo  {journal} {Nat Phys}\ }\textbf {\bibinfo
  {volume} {7}},\ \bibinfo {pages} {459} (\bibinfo {year} {2011})}\BibitemShut
  {NoStop}%
\bibitem [{\citenamefont {Falk}\ \emph {et~al.}(2014)\citenamefont {Falk},
  \citenamefont {Klimov}, \citenamefont {Buckley}, \citenamefont {Iv\'{a}dy},
  \citenamefont {Abrikosov}, \citenamefont {Calusine}, \citenamefont {Koehl},
  \citenamefont {Gali},\ and\ \citenamefont {Awschalom}}]{Falk2014}%
  \BibitemOpen
  \bibfield  {author} {\bibinfo {author} {\bibfnamefont {A.~L.}\ \bibnamefont
  {Falk}}, \bibinfo {author} {\bibfnamefont {P.~V.}\ \bibnamefont {Klimov}},
  \bibinfo {author} {\bibfnamefont {B.~B.}\ \bibnamefont {Buckley}}, \bibinfo
  {author} {\bibfnamefont {V.}~\bibnamefont {Iv\'{a}dy}}, \bibinfo {author}
  {\bibfnamefont {I.~A.}\ \bibnamefont {Abrikosov}}, \bibinfo {author}
  {\bibfnamefont {G.}~\bibnamefont {Calusine}}, \bibinfo {author}
  {\bibfnamefont {W.~F.}\ \bibnamefont {Koehl}}, \bibinfo {author}
  {\bibfnamefont {A.}~\bibnamefont {Gali}}, \ and\ \bibinfo {author}
  {\bibfnamefont {D.~D.}\ \bibnamefont {Awschalom}},\ }\href
  {http://link.aps.org/doi/10.1103/PhysRevLett.112.187601} {\bibfield
  {journal} {\bibinfo  {journal} {Physical Review Letters}\ }\textbf {\bibinfo
  {volume} {112}},\ \bibinfo {pages} {187601} (\bibinfo {year}
  {2014})}\BibitemShut {NoStop}%
\bibitem [{\citenamefont {Ajoy}\ and\ \citenamefont
  {Cappellaro}(2012)}]{Ajoy2012}%
  \BibitemOpen
  \bibfield  {author} {\bibinfo {author} {\bibfnamefont {A.}~\bibnamefont
  {Ajoy}}\ and\ \bibinfo {author} {\bibfnamefont {P.}~\bibnamefont
  {Cappellaro}},\ }\href {http://link.aps.org/doi/10.1103/PhysRevA.86.062104}
  {\bibfield  {journal} {\bibinfo  {journal} {Physical Review A}\ }\textbf
  {\bibinfo {volume} {86}},\ \bibinfo {pages} {62104} (\bibinfo {year}
  {2012})}\BibitemShut {NoStop}%
\bibitem [{\citenamefont {Bourgeois}\ \emph {et~al.}(2015)\citenamefont
  {Bourgeois}, \citenamefont {Jarmola}, \citenamefont {Gulka}, \citenamefont
  {Hruby}, \citenamefont {Budker},\ and\ \citenamefont
  {Nesladek}}]{2015arXiv150207551B}%
  \BibitemOpen
  \bibfield  {author} {\bibinfo {author} {\bibfnamefont {E.}~\bibnamefont
  {Bourgeois}}, \bibinfo {author} {\bibfnamefont {A.}~\bibnamefont {Jarmola}},
  \bibinfo {author} {\bibfnamefont {M.}~\bibnamefont {Gulka}}, \bibinfo
  {author} {\bibfnamefont {J.}~\bibnamefont {Hruby}}, \bibinfo {author}
  {\bibfnamefont {D.}~\bibnamefont {Budker}}, \ and\ \bibinfo {author}
  {\bibfnamefont {M.}~\bibnamefont {Nesladek}},\ }\href@noop {} {\bibfield
  {journal} {\bibinfo  {journal} {ArXiv e-prints}\ } (\bibinfo {year}
  {2015})},\ \Eprint {http://arxiv.org/abs/1502.07551} {arXiv:1502.07551
  [cond-mat.mes-hall]} \BibitemShut {NoStop}%
\bibitem [{\citenamefont {Cochrane}\ \emph {et~al.}(2012)\citenamefont
  {Cochrane}, \citenamefont {Lenahan},\ and\ \citenamefont
  {Lelis}}]{Cochrane2012}%
  \BibitemOpen
  \bibfield  {author} {\bibinfo {author} {\bibfnamefont {C.~J.}\ \bibnamefont
  {Cochrane}}, \bibinfo {author} {\bibfnamefont {P.~M.}\ \bibnamefont
  {Lenahan}}, \ and\ \bibinfo {author} {\bibfnamefont {A.~J.}\ \bibnamefont
  {Lelis}},\ }\href {http://dx.doi.org/10.1063/1.3675857} {\bibfield  {journal}
  {\bibinfo  {journal} {Applied Physics Letters}\ }\textbf {\bibinfo {volume}
  {100}},\ \bibinfo {pages} {23503} (\bibinfo {year} {2012})}\BibitemShut
  {NoStop}%
\bibitem [{\citenamefont {Lee}\ \emph {et~al.}(2012)\citenamefont {Lee},
  \citenamefont {Paik}, \citenamefont {McCamey},\ and\ \citenamefont
  {Boehme}}]{Lee2012}%
  \BibitemOpen
  \bibfield  {author} {\bibinfo {author} {\bibfnamefont {S.-Y.}\ \bibnamefont
  {Lee}}, \bibinfo {author} {\bibfnamefont {S.}~\bibnamefont {Paik}}, \bibinfo
  {author} {\bibfnamefont {D.~R.}\ \bibnamefont {McCamey}}, \ and\ \bibinfo
  {author} {\bibfnamefont {C.}~\bibnamefont {Boehme}},\ }\href {\doibase
  10.1103/PhysRevB.86.115204} {\bibfield  {journal} {\bibinfo  {journal}
  {Physical Review B}\ }\textbf {\bibinfo {volume} {86}},\ \bibinfo {pages}
  {115204} (\bibinfo {year} {2012})}\BibitemShut {NoStop}%
\bibitem [{\citenamefont {Lohrmann}\ \emph {et~al.}(2015)\citenamefont
  {Lohrmann}, \citenamefont {Iwamoto}, \citenamefont {Bodrog}, \citenamefont
  {Castelletto}, \citenamefont {Ohshima}, \citenamefont {Karle}, \citenamefont
  {Gali}, \citenamefont {Prawer}, \citenamefont {McCallum},\ and\ \citenamefont
  {Johnson}}]{Lohrmann2015}%
  \BibitemOpen
  \bibfield  {author} {\bibinfo {author} {\bibfnamefont {A.}~\bibnamefont
  {Lohrmann}}, \bibinfo {author} {\bibfnamefont {N.}~\bibnamefont {Iwamoto}},
  \bibinfo {author} {\bibfnamefont {Z.}~\bibnamefont {Bodrog}}, \bibinfo
  {author} {\bibfnamefont {S.}~\bibnamefont {Castelletto}}, \bibinfo {author}
  {\bibfnamefont {T.}~\bibnamefont {Ohshima}}, \bibinfo {author} {\bibfnamefont
  {T.~J.}\ \bibnamefont {Karle}}, \bibinfo {author} {\bibfnamefont
  {A.}~\bibnamefont {Gali}}, \bibinfo {author} {\bibfnamefont {S.}~\bibnamefont
  {Prawer}}, \bibinfo {author} {\bibfnamefont {J.~C.}\ \bibnamefont
  {McCallum}}, \ and\ \bibinfo {author} {\bibfnamefont {B.~C.}\ \bibnamefont
  {Johnson}},\ }\href {http://dx.doi.org/10.1038/ncomms8783 10.1038/ncomms8783}
  {\bibfield  {journal} {\bibinfo  {journal} {Nat Commun}\ }\textbf {\bibinfo
  {volume} {6}} (\bibinfo {year} {2015})},\ \Eprint
  {http://arxiv.org/abs/1503.07566} {arXiv:1503.07566 [cond-mat.mtrl-sci]}
  \BibitemShut {NoStop}%
\end{thebibliography}

%


\end{document}